%% file: LK_polar_quantum_v3.tex
\documentclass[prl,aps,a4paper,english,11pt,nofootinbib,notitlepage,superscriptaddress,onecolumn,accepted=2019-07-29]{quantumarticle}
\usepackage{import}
\input{macros}\usepackage{graphicx}
\usepackage{caption}
\usepackage{array}
\newcolumntype{L}[1]{>{\raggedright\let\newline\\\arraybackslash\hspace{0pt}}m{#1}}
\newcolumntype{C}[1]{>{\centering\let\newline\\\arraybackslash\hspace{0pt}}m{#1}}
\newcolumntype{R}[1]{>{\raggedleft\let\newline\\\arraybackslash\hspace{0pt}}m{#1}}
\usepackage{subcaption}
\begin{document}
\title{
	\hbox{A polar decomposition for quantum channels}\vspace{2mm}
\hbox{(with applications to bounding error propagation in} quantum circuits)
}
\author{Arnaud Carignan-Dugas}
\email{arnaud.carignan@gmail.com}
\orcid{0000-0002-2036-2688}
\affiliation{Institute for Quantum Computing and the Department of Applied
	Mathematics, University of Waterloo, Waterloo, Ontario N2L 3G1, Canada}
\author{Matthew Alexander}
\affiliation{Institute for Quantum Computing and the Department of Applied
	Mathematics, University of Waterloo, Waterloo, Ontario N2L 3G1, Canada}
\author{Joseph Emerson}
\affiliation{Institute for Quantum Computing and the Department of Applied
	Mathematics, University of Waterloo, Waterloo, Ontario N2L 3G1, Canada}
\affiliation{Canadian Institute for Advanced Research, Toronto, Ontario M5G 1Z8, Canada}

\maketitle

\begin{abstract}
	Inevitably, assessing the overall performance of a quantum computer must rely on characterizing some of its elementary constituents and, from this information,
	formulate a broader statement concerning more complex constructions thereof. 
	However, given the vastitude of possible quantum errors as well as their coherent nature,
	accurately inferring the quality of composite operations
	is generally difficult. 
	To navigate through this jumble,
	we introduce a non-physical 
	simplification of quantum maps that we refer to as the leading Kraus (LK) approximation. The uncluttered parameterization of LK approximated maps naturally suggests the introduction of a unitary-decoherent polar factorization for quantum channels in any dimension. 
	We then leverage this structural dichotomy to bound the evolution -- as circuits grow in depth -- of two of the most experimentally relevant figures of merit, namely the average process fidelity and the unitarity. We demonstrate that 
	the leeway in the behavior of the process fidelity is essentially taken into account by physical unitary operations.
\end{abstract}

\section{Introduction}
Just like evaluating a piano doesn't involve playing all possible pieces of 
music, characterizing a computer (classical or quantum) doesn't involve running all infinitely many 
circuits. The natural procedure to characterize both these devices is to 
gather information on a restricted number of components, and based on that
information make conclusions on the quality of more involved constructions (melodies, chords, circuits, magic state injections, etc).
When considering the tuning of a piano, the extrapolation is not much of a problem; imperfections are typically tied to specific keys, and they don't tend to propagate over the keyboard as the music goes on, and unless there is some resonant effect, the errors don't coherently interfere. Hence, the quality
of individual keys generally guarantees playability. 
In this sense, the characterization of a piano is similar to that of a classical computer: the well-behaved stochasticity of the noise eases
the passage between an assertion of components quality to a broader assertion on
the performance of more complex operations. This statement can be phrased the other way around: a limited range of behaviors simplifies the search for imperfections.

In contrast, when characterizing a quantum computer, the jump from a characterization of elementary operations to a quantified assertion on
the overall device performance is more knotty; errors can coherently interfere and propagate through the entire device via multi-qubit operations. 
This thorny situation can be quantified, for instance, by bounding the behavior of the average process fidelity (hereafter the fidelity and its counterpart, the infidelity), an experimentally important figure of merit which captures the overlap between an implemented operation and its target.
More precisely, one may ask: ``What are the best and worst fidelities of a circuit given a knowledge of the fidelity of its components?'' When dealing with a classical scenario, we would expect the difference between the best and worst cases to remain insignificant (remember the piano analogy). In a quantum scenario, however, it is known that the largest discrepancy, which is achieved by unitary errors, grows quickly (quadratically) in the circuit depth (see, for instance, Carignan-Dugas et al \cite{Dugas2016}). Not so surprisingly, the best case corresponds to a unitary cancellation, and the worst case corresponds to a coherent buildup. 
This lead to another question: ``What if we are guaranteed that the individual 
errors are not unitary?'' In particular, what if we measure the degree to which the error operations are unitary, known as the unitarity\cite{Wallman2015}, an experimental figure of merit which captures the coherence in the noise? Previous work has given partial answers to this question: Carignan-Dugas et al  \cite{Dugas2016} derive bounds that fall back to the ``piano analogy'' when the unitarity is minimal; additionally, they provide examples of quantum
channels that saturate their bound in the intermediate regime where errors 
are neither purely unitary nor purely stochastic, but still unital and acting on 
a single qubit coupled with a system of arbitrary (but finite) size.\footnote{They attribute all the error dynamics on the qubit; the intuitive geometric picture offered by parameterization of processes acting 
	on the Bloch sphere allows showing the saturation of the bound for unital channels. 
	The bound in the non-unital case included a dimensional factor which prevented its saturation.}.
In this paper, we generalize that bound to all dimensions and show its near saturation (i.e. to second order in the infidelity or better) and also account for non-unital processes. That is, we provide a closely saturated bound for all finite-dimensional quantum channels.
While this is already an interesting result, the tools that we develop to generalize the bound help us answering a far more fundamental question. In previous work, the saturation was shown through a handful of examples. Now, we provide a complete descriptive answer to:
\begin{center}
	\emph{What is the set of mechanisms responsible for the discrepancy between the best and the worst fidelity of a circuit?}
\end{center}
This would not be much of a fundamental question if the answer didn't also unravel an important dichotomy in classifying quantum errors. Given the intricate geometry of quantum states \cite{GeoQuantStates}, the answer could have included some obscure blend of non-intuitive mechanisms, leaving us with yet another resignation in the attempt to intuitively reason about quantum dynamics. Although, for once, this is not the case: the discrepancy between
the best and worst fidelity is, to high precision, entirely taken into account
by unitary dynamics\footnote{Given realistic errors, which are properly defined in \cref{sec:dephasers}, and are more formally referred to as ``equable''. The equability assumption corresponds to ruling out two types of errors. 1) Extreme dephasing effects between a small set of states and the rest of the systems. 2) Extreme Hamiltonian alterations.}. Even more surprisingly, the unitary dynamics itself is precisely the product of the ``unitary factors'' of individual circuit components. As we demonstrate through \cref{thm:decomp}, every non-catastrophic channel (see \cref{defn:nearly_unitary}) can be decomposed as a physical unitary
followed or preceded by a decoherent channel. For realistic errors, the unitary is unique and is referred to as the coherent factor. This factorization is analogous to the well-known matrix polar decomposition and, as we will show, directly stems from it. The uniqueness of the coherent factor might puzzle the skeptical reader. For example, how should we unambiguously define such factor in the case of an error which consists of a mixture of near-identity unitaries (i.e. $\mc A(\rho) = \sum_i p_i U_i \rho U_i^\dagger$, where $U_i \approx \mbb I_d$)? Should it be the unitary operation with the highest weight? Should it relate with some kind of ensemble average over the associated Hamiltonians? To systematically answer this type of question, we introduce
the leading Kraus (LK) approximation (see \cref{defn:lka}), a sub-parameterization of quantum channels
which, among other things, exposes a natural definition for the coherent and decoherent factors of a channel. 

What allows us to really profit from the channel polar decomposition is the surprising property that the LK approximation, despite its seemingly bare structure, closely captures the evolution of the fidelity and unitarity in circuits. That is, we can mathematically replace all the channels in a circuit by their respective LK approximation and still expect to accurately bound its fidelity and unitarity (see \cref{thm:uni_evo,thm:fid_evo}). Working with the uncluttered structure offered by the LK approximation helped us identify and rule out pathological error scenarios, which we refer to as ``extremal'' (see \cref{sec:dephasers} for more details). For all realistic noisy channels, we derive the following observations (they hold to high precision):
\begin{enumerate}[i.]
	\item The infidelity (the counterpart to the fidelity) of a channel
	can be split into two terms (see \cref{thm:fid_evo_3} and the discussion that immediately follows): 
	\begin{enumerate}
		\item a coherent infidelity, which corresponds to the infidelity of the coherent factor to the target channel;
		\item a decoherent infidelity, which corresponds to the infidelity of the decoherent factor to the identity.
	\end{enumerate}
	\item The decoherent infidelity of a channel is in one-to-one correspondence with its unitarity. Moreover, the decoherent infidelity corresponds to the minimum infidelity of the channel after the application of a unitary (the coherent infidelity is correctable through a composition with a unitary). (See \cref{thm:max_cor}.)
	\item The unitarity of a composite channel is a decay function expressed in terms of individual channels' unitarity. (See \cref{thm:uni_evo_2}.)
	\item The fidelity of the composition of decoherent channels is a decay function expressed in terms of individual channels' fidelity. (See \cref{thm:fid_evo_2}.)
	\item The fidelity of a general composition is upper bounded by a decay dictated by the decoherent factors (hence by the unitarity of individual components). (See \cref{thm:max_cor_multi}.)
	\item The discrepancy between the upper and the lower bound of the fidelity is captured by the fidelity of the composition of the coherent factors (to the target circuit). (See \cref{thm:fid_evo_3}.)
\end{enumerate}
These realizations are directly applicable to the analysis and development of process characterization methods. The fidelity of various error processes
can be robustly and efficiently estimated through a scalable experimental protocol known as randomized benchmarking (RB) \cite{Emerson2005,Dankert2009,Magesan2011,Magesan2012a} and a family of generalizations thereof \cite{Knill2008,Magesan2012b,GambettaCorcoles2012,Gaebler2012,Granade2014,Barends2014,Wallman2014,Dugas2015,Wallman2015,Wallman2015b,Sheldon2016,Cross2016,Combes2017,Hashagen2018,Brown2018,Franca2018,Helsen2018,Proctor2018}. 
To remain efficient as quantum devices grow larger, RB experiments only extract partial information about specific sets of components. 
A known challenge is to
leverage this limited view to formulate a more rounded understanding of the device.
By looking at the fidelity of well-designed compositions, it should be possible to extract other 
figures of merit attached to quantum processes. The idea is that since process matrices dictate 
the evolution of the fidelity, conversely, the evolution of the fidelity can tell us information 
about process matrices. However, given the generally large amount of parameters involved in process matrices, it is not always immediately clear how the signal obtained from extracting the fidelity of various circuit compositions connects with quantities of interest. The above six enumerated observations allow to make more sense out of such signals. 

We structure the paper as follows. In \cref{sec:chan_props}, we introduce
important characterization figures of merit -- the average process fidelity
and the unitarity -- and relate them with the Kraus operator formalism. In
\cref{sec:evo}, we define the LK approximation and present its aptitude
in capturing important characteristics of evolving quantum circuits. 
In  \cref{sec:polar}, based on the emergent mathematical structure 
of LK approximated channels, we 
show the existence of a channel polar unitary-decoherent decomposition.
In \cref{sec:assesment}, we make use of the approximation to demonstrate key behavioral aspects of quantum circuits based on partial knowledge of their components.

For the sake of conciseness, most demonstrations are pushed to the appendix. Moreover, in the main text,
certain results have been abridged by gathering 
higher order terms under the acronym ``H.O.T.''.
The complete expressions --
which are not any more insightful than their abbreviated analog -- 
are provided in the appendix.

\section{Channel properties captured by the leading Kraus operator}\label{sec:chan_props}
A quantum channel is a completely-positive (CP), trace-preserving (TP) map acting on $M_d(\mbb C)$. 
Given a quantum channel $\mc A: M_d(\mbb C) \rightarrow M_d (\mbb C)$, the Choi matrix 
of $\mc A$ is defined as \cite{Choi1975}
\begin{align}
{\rm Choi}(\mc A) := \sum_{ij} E_{ij} \otimes \mc A( E_{ij})~, \tag {Choi matrix}
\end{align}
where
\begin{align}
E_{ij} := e_i e_j^{\dagger}~,
\end{align}
and $e_i$ are canonical orthonormal vectors.
The Choi matrix is positive semi-definite iff $\mc A$ is CP, and has 
trace $d$ if $\mc A$ is TP or unital\footnote{A channel $\mc A$ is unital iff $\mc A(\mbb I_d) = \mbb I_d $.}. Since ${\rm Choi}(\mc A) \geq 0$, it has a spectral
decomposition of the form 
\begin{align}
{\rm Choi}(\mc A) &:= \sum_{i=1}^{d^2}  {\rm col}(A_i) {\rm col}^\dagger(A_i)~, \label{eq:spectral_Choi}\\
& = \sum_{i=1}^{d^2} \|A_i\|_2^2 {\rm col}(\nlz{A}_i) {\rm col}^\dagger(\nlz{A}_i)~, 
\end{align}
where ${\rm col}(A) \in \mbb C^{d^2}$ denotes the column vectorization of a matrix $A \in M_d(\mbb C)$\footnote{$    {\rm col} (A) := \sum_{ij}A_{ij} e_j \otimes e_i$}, $\|\cdot\|_p$ denotes the Schatten $p$-norm, and $\nlz{A} = A/ \|A\|_2$ denotes normalized matrices with respect to the Schatten $2$-norm. The eigenvectors ${\rm col}({\nlz{A}_i})$ are orthonormal, an without loss of generality the eigenvalues are ordered with respect to the Frobenius norm (Schatten 2-norm):
\begin{align}
\|A_1\|_2^2 \geq \|A_2\|_2^2 \geq \cdots \geq \|A_{d^2}\|_2^2 \geq 0~.
\end{align}
Given a spectral decomposition like \cref{eq:spectral_Choi}, we
can express the channel's action on states $\rho \in M_d(\mbb C)$ as \cite{Kraus1983states}:
\begin{align}
\mc A(\rho) &=\sum\limits_{i=1}^{d^2} A_i \rho A_i^\dagger~, \tag{Kraus decomposition}
\end{align}
with\begin{align}
\langle A_i , A_j \rangle = \|A_i\|_2^2 \delta_{ij}~,
\end{align}
where the usual Hilbert-Schmidt inner product is used.
Notice that the TP condition implies that $\sum_i (\|A_i\|_2^2/d)=1$.
The matrices $A_i\in M_d(\mbb C)$ are referred to as (ordered) canonical Kraus operators. In this work, $A_1$ (which is associated with the highest Choi matrix eigenvalue $\|A_1\|_2^2$) will deserve special attention, 
and is attributed the title of ``leading Kraus (LK) operator''. 
In general, $A_1$ might be non-unique when the spectrum of the Choi matrix is 
degenerate. However, in this work we focus on non-catastrophic channels 
(\cref{defn:nearly_unitary}), for which $A_1$ is unique. 

Given an operation $\mc A$ and a target unitary channel $\mc U(\rho) = U \rho U^\dagger$ \footnote{For unitaries, we used the calligraphic font to denote the channel and the non-calligraphic one to denote its associated $d \times d$ unitary matrix.}, we can compare the overlap of their outputs given specific inputs $M \in M_d(\mbb C)$ through the $M$-fidelity:
\begin{align}
f_M (\mc A, \mc U):= \frac{\inner{\mc A (M)}{\mc U (M)}}{\|M\|_2^2}~.
\end{align}
The well-known average gate fidelity is obtained by averaging the $M$-fidelities uniformly (i.e. with respect to the Haar measure) over
all physical pure states $|\psi\rangle \langle \psi|$:
\begin{align}
F(\mc A, \mc U):=\mbb E_{\text{Haar}} ~f_{|\psi\rangle \langle \psi|}(\mc A, \mc U) ~. \label{def:fid}
\end{align}
The average infidelity $r$ is simply a shorthand for $1-F$.  
Instead of averaging over quantum states, we could also average uniformly over all 
operators $M \in M_d(\mbb C)$. More precisely, given any orthogonal 
operator basis $\{B_i\}$ for $M_d(\mbb C)$, we can uniformly average over the 
$M$-fidelities $f_{B_i}$, which yields the average process fidelity \footnote{For the readers familiar with the $\chi$-matrix, $\Phi(\mc A, \mc U)$ is a way to express the well-known $\chi_{00}$ element. Of course, the $\chi$-matrix has to be defined with respect to an orthonormal operator basis $\{B_i\}$ with $B_0=U$. Some might also be more familiar with the notion of entanglement fidelity, which is again $\Phi$.}
\begin{align}\label{eq:phi_average}
\Phi(\mc A, \mc U):= \mbb E_{\{B_i\}}~ f_{B_i}(\mc A, \mc U)~.
\end{align}
Compared to $\Phi$, $F$
puts a slightly higher weight over the identity component $\mbb I_d$. The TP condition enforces this special component to take a fixed value, $f_{\mbb I_d}=1$. Hence the two quantities are closely related via \cite{Nielsen2002}:
\begin{align}\label{eq:fid}
F(\mc A, \mc U) = \frac{d \Phi(\mc A, \mc U)+1}{d+1}~.
\end{align}
$F(\mc A, \mc U)$ is the overlap between the output state $\mc A(\rho)$
of an implemented channel $\mc A$ and its ideal output $\mc U (\rho)$, averaged over all physical pure input states $|\psi\rangle \langle \psi|$.
While $F(\mc A, \mc U)$ conveys a more graspable interpretation, it will remain easier here to 
work with $\Phi(\mc A, \mc U)$ since it ties with the Kraus operators through
\begin{align}\label{eq:chi00_expr}
\Phi (\mc A, \mc U) = \sum\limits_{i=1}^{d^2} \left|\left\langle \frac{A_i}{\sqrt{d}} ,\frac{U}{\sqrt{d}}\right\rangle \right|^2 = \sum\limits_{i=1}^{d^2} (\|A_i\|^2_2/d) \abs{\inner{\nlz{A}_i}{ U / \sqrt{d}}}^2~.
\end{align}
Since $\{\nlz{A}_i\}$ forms an orthonormal basis and $\|U/\sqrt{d}\|_2=1$, it follows that
\begin{align}
\sum\limits_{i=1}^{d^2} \abs{\inner{\nlz{A}_i}{ U / \sqrt{d}}}^2=1~.
\end{align}
If $\|A_i\|^2_2/d$ can be thought as the ``weights'' of the Kraus operators, 
$\abs{\inner{\nlz{A}_i}{ U / \sqrt{d}}}^2$ can be thought as normalized overlaps
with the target $U$. 

To quantify the coherence of a quantum channel, 
one could wonder how much the Bloch vectors (the traceless component of quantum states \cite{Bloch1946})
are contracted. For instance, consider the unitarity, which is the squared length ratio of the Bloch 
vectors before and after the action of the channel $\mc A$, averaged over all
physical Bloch vector inputs corresponding to pure states $|\psi\rangle \langle \psi|-\mbb I_d/d$ \cite{Wallman2015}:
\begin{align}
u(\mc A):= \mbb E_{\text{Haar}}~  \frac{\|\mc A(|\psi\rangle \langle \psi|-\mbb I_d/d)\|_2^2}{\||\psi\rangle \langle \psi|-\mbb I_d/d\|_2^2}~.
\end{align}
Let's extend the domain of $\Phi$ to include a new function of $\mc A$:
\begin{align}\label{eq:calc_uni}
\Upsilon(\mc A):= \sqrt{\Phi(\mc A^\dagger \mc A, \mc I)} =  \sqrt{\sum\limits_{i,j=1}^{d^2} \left|\left\langle \frac{A_j^\dagger A_i}{\sqrt{d}} ,\frac{I}{\sqrt{d}}\right\rangle \right|^2}  = \sqrt{ \sum\limits_{i=1}^{d^2}  \left(\frac{\|A_i\|_2^2}{d}\right)^2} ~.
\end{align}
Straightforward calculations closely relate the unitarity to $\Upsilon$ via
\begin{align}
u(\mc A)=\frac{d^2 \Upsilon^2(\mc A)-1}{d^2-1}. \label{eq:chi_2_uni}
\end{align}
(Notice that the notation alludes to the connection between greek and latin alphabets; it relates ``phi'' to ``F'' and ``upsilon'' to ``u''.)

We are ready to express a first result:
\begin{boxlem}{}{uni}
	Consider a CPTP map $\mc A$ with ordered canonical Kraus decomposition
	\begin{align}
	\mc A(\rho) =\sum\limits_{i=1}^{d^2} A_i \rho A_i^\dagger~. \notag
	\end{align}
	Then, 
	\begin{align}
	0 \leq \Upsilon^2(\mc A) - \left(\frac{\|A_1\|_2^2}{d}\right)^2\leq  \left(1-  \Upsilon^2(\mc A) \right)^2~. \label{eq:uni}
	\end{align}    
\end{boxlem}

\begin{proof}
	$ \Upsilon^2(\mc A)$ can be expanded as a sum over $d^2$ terms:
	\begin{align}
	\Upsilon^2(\mc A) = & 
	\sum \limits_{\substack{i}} \left(\frac{\|A_{i}\|_2^2}{ d}\right)^2~. \label{eq:unitarity_single_expansion}
	\end{align}
	Using H\"older's inequality on the RHS yields
	\begin{align}
	\Upsilon^2(\mc A) \leq \max_i\frac{\|A_{i}\|_2^2}{ d} \sum \limits_{\substack{j}} \frac{\|A_{j}\|_2^2}{ d}= \|A_1\|_2^2/d~. \label{eq:lower_bound_A_1_norm}
	\end{align}
	Using this lower bound on $\|A_1\|_2$, we get
	\begin{align}
	\Upsilon^2(\mc A) &=  \left(\frac{\|A_{1}\|_2^2}{ d}\right)^2+\sum \limits_{\substack{i \neq 1}} \left(\frac{\|A_{i}\|_2^2}{ d}\right)^2 \\
	&\leq \left(\frac{\|A_{1}\|_2^2}{ d}\right)^2+ \left(\sum \limits_{\substack{i \neq 1}} \frac{\|A_{i}\|_2^2}{ d}\right)^2~\notag \\
	& = \left(\frac{\|A_{1}\|_2^2}{ d}\right)^2+ \left(1-    \frac{\|A_{1}\|_2^2}{d}\right)^2 \tag{$\sum \limits_{\substack{i}} \frac{\|A_{i}\|_2^2}{ d}=1$}\\
	& \leq \left(\frac{\|A_{1}\|_2^2}{ d}\right)^2+  \left(1- \Upsilon^2(\mc A)\right)^2  \tag{\cref{eq:lower_bound_A_1_norm}}
	\end{align}
	From \cref{eq:unitarity_single_expansion}, it follows that
	$\left(\frac{\|A_{1}\|_2^2}{ d}\right)^2 \leq \Upsilon^2(\mc A)$, which completes the proof.
\end{proof}
It follows from \cref{eq:uni} that $\Upsilon^2(\mc A) > 1/2$
is a sufficient condition to guarantee the uniqueness of $A_1$\footnote{Indeed, it implies that $\|A_1\|_2^2/d > 1/\sqrt{2}>1/2$.}.  
This partially motivates the following definition:
\begin{boxdefn}{non-catastrophic channels}{nearly_unitary}
	A channel $\mc A$ is said to be non-catastrophic if it overlaps
	enough with its targeted unitary channel $\mc U$:
	\begin{align}
	\Phi (\mc A, \mc U) & > 1/2~,\label{eq:nc_cond1}
	\end{align}
	and if it doesn't greatly contract the Bloch vectors:
	\begin{align}
	\Upsilon^2(\mc A) & > 1/2~.\label{eq:nc_cond2}
	\end{align}    
\end{boxdefn}
The condition described by \cref{eq:nc_cond1} allows us to express our second result:
\begin{boxlem}{}{fid}
	Consider a non-catastrophic channel $\mc A$ with unitary target $\mc U$ and ordered canonical Kraus decomposition
	\begin{align}
	\mc A(\rho) =\sum\limits_{i=1}^{d^2} A_i \rho A_i^\dagger~. \notag 
	\end{align}
	Then,
	\begin{align}
	0 \leq \Phi(\mc A, \mc U)- \abs{\inner{\frac{A_1}{\sqrt{d}}}{\frac{U}{\sqrt{d}}}}^2  \leq (1-\Upsilon^2(\mc A))(1-\Phi(\mc A, \mc U))~.
	\end{align}    
\end{boxlem}
\begin{proof}
	
	Using H\"older's inequality on the RHS of \cref{eq:chi00_expr}, we have
	\begin{align}
	\Phi(\mc A, \mc U) &\leq \max\limits_{i} \abs{\inner{\nlz{A}_i}{ \nlz{U}}}^2  \sum\limits_{j=1}^{d^2} (\|A_j\|^2_2/d) = 
	\max\limits_{i} \abs{\inner{\nlz{A}_i}{ \nlz{U}}}^2 ~.\label{eq:lower_bounds_prob_inner}
	\end{align}
	For non-catastrophic channels, it must be that $\max\limits_{i} \abs{\inner{\nlz{A}_i}{ \nlz{U}}}^2 =  \abs{\inner{\nlz{A}_1}{ \nlz{U}}}^2$.
	To see this more clearly, let $ \abs{\inner{\nlz{A}_1}{ \nlz{U}}}^2 = 1/2-\epsilon_1$
	and $\|A_1\|_2^2/d = 1/2 +\epsilon_2$, where $\epsilon_2>0$ from the non-catastrophic condition. Then, consider the following inequalities:
	\begin{align}
	\Phi(\mc A, \mc U) &=  \abs{\inner{\nlz{A}_1}{ \nlz{U}}}^2 (\|A_1\|_2^2/d) + \sum_{i \neq 1} \abs{\inner{\nlz{A}_i}{ \nlz{U}}}^2 (\|A_i\|_2^2/d)  \tag{\Cref{eq:chi00_expr}} \\
	& \leq \abs{\inner{\nlz{A}_1}{ \nlz{U}}}^2 (\|A_1\|_2^2/d) +  \left(\sum\limits_{i \neq 1} (\|A_i\|_2^2 / d)\right) \left(\sum\limits_{j \neq 1} \abs{\inner{\nlz{A}_j}{ \nlz{U}}}^2 \right)  \notag \\
	& = \abs{\inner{\nlz{A}_1}{ \nlz{U}}}^2 (\|A_1\|_2^2/d) + (1-\abs{\inner{\nlz{A}_1}{ \nlz{U}}}^2)(1-\|A_1\|_2^2/d)  \label{eq:split_sum}\\
	& = 1/2 -2\epsilon_1 \epsilon_2~.
	\end{align}
	From the non-catastrophic condition, $\epsilon_1 <0$, which implies that $\abs{\inner{\nlz{A}_1}{ \nlz{U}}}^2>1/2$.
	
	Hence, \cref{eq:lower_bounds_prob_inner} can
	be reexpressed into $1-\abs{\inner{\nlz{A}_1}{ \nlz{U}}}^2 \leq 1-\Phi(\mc A, \mc U)$,
	which yields the following:
	\begin{align}
	\Phi(\mc A, \mc U) &\leq  \abs{\inner{\nlz{A}_1}{ \nlz{U}}}^2 (\|A_1\|_2^2/d) + (1-\abs{\inner{\nlz{A}_1}{ \nlz{U}}}^2)(1-\|A_1\|_2^2/d) \tag{\Cref{eq:split_sum}} \\
	& \leq  (\|A_1\|_2^2 / d)\abs{\inner{\nlz{A}_1}{ \nlz{U}}}^2 + (1-\Upsilon^2(\mc A))(1-\Phi(\mc A, \mc U))~. \tag{\cref{eq:lower_bound_A_1_norm}}
	\end{align}
	From \cref{eq:chi00_expr} we also have $(\|A_1\|_2^2 / d)\abs{\inner{\nlz{A}_1}{ \nlz{U}}}^2 \leq \Phi(\mc A, \mc U)$, which completes the proof.
\end{proof}

The LK operator alone provides a very accurate
approximation of $1-\Phi$ and $1-\Upsilon$.
This only begins a list of realizations regarding the role of
LK operators in quantum dynamics. As we will see, they 
also contain most of the information necessary to describe the
\emph{evolution} of $\Phi$ and $\Upsilon$.
\section{The LK approximation and two evolution theorems}\label{sec:evo}

The last section naturally suggests the following channel approximation as a 
means to partially characterize non-catastrophic quantum dynamics:

\begin{boxdefn}{the Leading Kraus (LK) approximation}{lka}
	Consider a channel $\mc A: M_d(\mbb C) \rightarrow M_d(\mbb C)$ with leading Kraus operator $A_1$. We define its leading Kraus (LK) approximation as:
		\begin{align}
	{\mc A}^\star(\rho)= A_1\rho A_1^\dagger~.
	\end{align}
\end{boxdefn}
 Notice that ${\mc A}^\star$ is always CP (${\rm Choi}(\mc A^\star) \geq 0$), but is TP iff $\mc A$ is unitary. Hence, $\mc A^\star$ fails to be generally physical. However, as we will see, it closely describes the dynamics of certain physical quantities, so one may qualify this map as ``quasi-dynamical''. The 
 general 
 specification of a map acting on a $d$-dimensional quantum system requires roughly $d^4$
 parameters, and due to the intricate geometry of quantum states, the parameterization of its range of action is quite convoluted. In contrast, the LK 
 approximation is remarkably transparent: it is fully parameterized by
 $d \times d$ matrices with spectral radius smaller than $1$ (contractions)
 and Frobenius norm greater than $d/\sqrt{2}$ \footnote{This last constraint only prevents catastrophic noise scenarios.}.
 If the noise is non-catastrophic, every quantum map has a
 corresponding LK approximation, and every $d\times d$ linear contraction
 corresponds to at least one quantum operator.
 
Given $m$ channels $\mc A_i$, we denote the composition \mbox{$\mc A_m \circ \mc A_{m-1} \circ \cdots \circ \mc A_2 \circ \mc A_1$} as
$\mc A_{m:1}$. Replacing every element of the composition by its LK approximation, {$\mc A_m^\star \circ \mc A_{m-1}^\star \circ \cdots \circ \mc A_2^\star \circ \mc A_1^\star$}, is noted as
$\mc A_{m:1}^\star$.  In general, the composition
operation doesn't commute with the LK approximation, that is $\mc A_{m:1}^\star \neq (\mc A_{m:1})^\star$. To put it in other words, the LK operator of 
a circuit is generally not the multiplication of the LK operators of its elements.
However, while $\mc A_{m:1}^\star$ provides an incomplete description of $\mc A_{m:1}$, they still might share some
comparable characteristics. That is, there might exist some function $f: \text{CP maps} \rightarrow \mbb R $
for which $f(\mc A_{m:1}^\star) \approx f(\mc A_{m:1})$. As we show,
not only there exist such functions, but some of them correspond to important experimental figures of merit. From the previous section, we know that \mbox{$\Phi(\mc A, \mc U) \approx \Phi(\mc A^\star, \mc U)$} and $\Upsilon(\mc A) \approx \Upsilon (\mc A^\star)$. What may be more surprising are  the following two theorems:
\begin{boxthm}{the unitarity of a circuit after approximating its elements}{uni_evo}
	Consider $m$ non-catastrophic channels $\mc A_i$ with respective unitary targets $\mc U_i$ 
	and suppose that the composition $\mc A_{m:1}$ is also non-catastrophic.
	Then,
	\begin{align}
	0 \leq  \Upsilon^2(\mc A_{m:1})-\Upsilon^2(\mc A_{m:1}^\star)
	&\leq (1- \Upsilon(\mc A_{m:1}^\star))^2 \leq (1- \Upsilon^2(\mc A_{m:1}))^2~. \label{eq:bounduni}
	\end{align}
\end{boxthm}

\begin{boxthm}{the fidelity of a circuit after approximating its elements}{fid_evo}
	Consider $m$ non-catastrophic channels $\mc A_i$ with respective unitary targets $ \mc U_i$ 
	and suppose that the composition $\mc A_{m:1}$ is also non-catastrophic.
	Then,
	\begin{align}
	&0 \leq \Phi(\mc A_{m:1}, \mc U_{m:1}) -\Phi(\mc A^\star_{m:1}, \mc U_{m:1})
	 <  \notag \\
	 &(1- \Phi(\mc A_{m:1}^\star, \mc U_{m:1}))\sum \limits_{\substack{i=1} }^{m}\left(1- \Upsilon(\mc A_i^\star)\right) \notag 
	 +\frac{1}{2} \left(\sum\limits_{i=1}^m (1-\Upsilon^\star(\mc A_i))\right)^2 \leq
	 \label{eq:boundfidstar} \\
	&(1- \Phi(\mc A_{m:1}, \mc U_{m:1}))\sum \limits_{\substack{i=1} }^{m}\left(1- \Upsilon^2(\mc A_i)\right)+	\frac{1}{2} \left(\sum\limits_{i=1}^m (1-\Upsilon^2(\mc A_i))\right)^2
	+\text{H.O.T.}~
	\end{align}
\end{boxthm}

$\mc A^\star$ differs from 
the veritable channel $\mc A$ in many ways as shown by comparing   
various $M$-fidelities $f_M(\mc A_{m:1}, \mc U_{m:1})$ with $f_M(\mc A_{m:1}^\star, \mc U_{m:1})$ (see two animated examples at \href{https://youtu.be/lTrBTIJHJJM}{https://youtu.be/lTrBTIJHJJM} and \href{https://youtu.be/A6i-k6eHsGM}{https://youtu.be/A6i-k6eHsGM}). Of course, some 
kind of 
discrepancy is expected since the LK approximation contains only $d^2$ parameters instead of $\sim d^4$. 
Essentially, the LK operators 
closely dictate the evolution of the average of $M$-fidelities $\Phi= \mbb E f_M$ (see \cref{eq:phi_average}), while 
the other Kraus operators add or subtract to specific $M$-fidelities 
$f_M$ in such a way that the sum of those variations almost exactly cancels.

The evolution theorems presented in this section will greatly help classify
different types of errors\footnote{An error channel simply refers to a channel with identity target $\mc I$.}. Indeed, they allow tying
behavioral signatures in the evolution of $\Upsilon$ and $\Phi$ to more digestible error profiles. In particular, the two theorems further motivate, as shown in \cref{sec:assesment}, the definition of a natural dichotomy in quantum channels (itself introduced in \cref{sec:polar}).

\section{A polar decomposition for quantum channels}\label{sec:polar}
\subsection{Defining decoherence}
\begin{figure}[!b]
\floatbox[{\capbeside\thisfloatsetup{capbesideposition={right,center},capbesidewidth=0.5\linewidth}}]{figure}[\FBwidth]
{\caption{Representation of the spiraling action of a normal matrix acting on a $2 \times 2$ subspace. The polar decomposition, in this case, separates the azimuthal and radial components of the action. Quantum dynamics on $d>2$ can generate spiraling actions on the Bloch space for which the rotation factor can't be interpreted as a physical unitary operation. In this sense, spiraling, despite generating some rotating action, is inherent to some decoherent dynamics.}\label{fig:spiral}}
{\includegraphics[width=6cm]{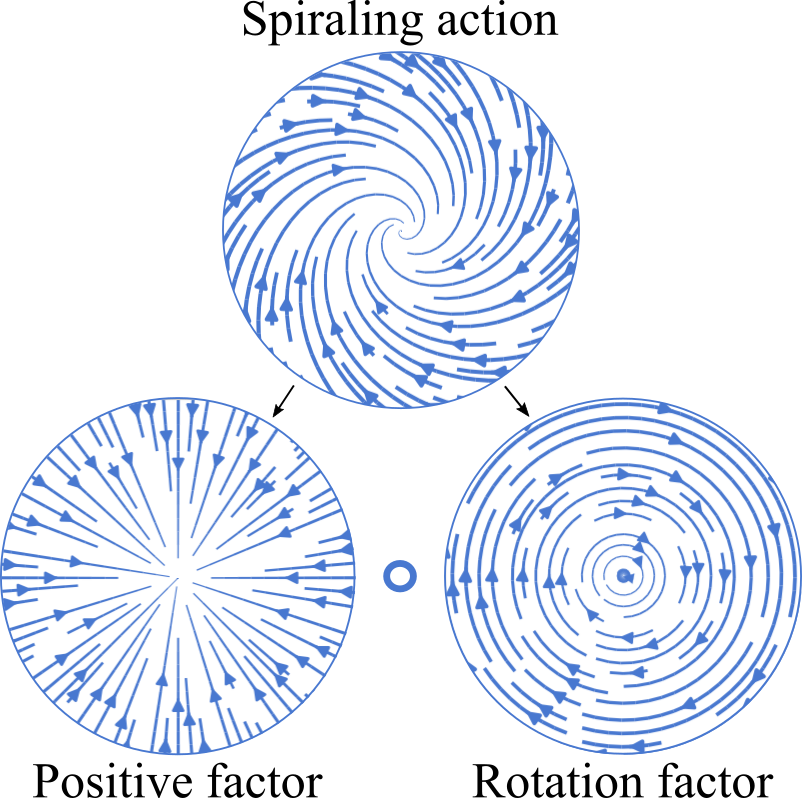}}
\end{figure}

Due to the intricate geometry of $d$-dimensional quantum states \cite{GeoQuantStates}, quantum processes can be
delicate to dissect. 
One of the main reasons the single qubit Bloch sphere is frequently invoked stems from the 
simple picture it offers:
\begin{enumerate}[i.]
	\item There is a clear bijection between quantum states and the Bloch ball \cite{Bloch1946}.
	\item The action on the Bloch vectors can be decomposed into a positive semi-definite contraction $|M| \leq \mbb I_{3}$, followed by orthogonal matrix $R \in O(3)$, which corresponds to a physical unitary $U \in SU(2)$, added to a translational vector $\vec{t}$ (the non-unital vector) \cite{Fujiwara1999,Ruskai2002,Bourbon2004}:
	\begin{align}\label{eq:bloch_action}
	\vec{v} \rightarrow \underbrace{R|M|}_{M}~  \vec{v}~+~\vec{t}~,
	\end{align}
	where $|M|$ denotes $(M^\dagger M)^{\frac{1}{2}}$. $M= R|M|$ is referred to the unital matrix. 
\end{enumerate} 
Not every contraction $|M|$ is physical; for instance, 
transforming the Bloch sphere into a disk violates CP-ness (the folkloric ``no pancake'' theorem \cite{RBK2010}). A thorough 
analysis of CPTP maps acting on $M_2(\mbb C)$ is provided in \cite{Ruskai2002}. 
For higher dimensions, the Bloch sphere imagery falls apart in many ways:
\begin{enumerate}[i.]
	\item The generalized Bloch space is not a {$(d^2-1)$}-ball (with 
	respect to the $2$-norm on $\mbb R^{d^2-1}$) \cite{GeoQuantStates}.
	\item If we express the action on the Bloch vector as in \cref{eq:bloch_action}
	where $R \in O(d^2-1)$ and $|M| \geq 0$, we realize that
	\begin{enumerate}
		\item $R$ generally doesn't correspond to a physical unitary operation in $SU(d)$
		(the unitary map defined by $\vec{v} \rightarrow R \vec{v}$ is not necessarily CP). 
		\item $|M|$ is not necessarily a contraction. Its spectrum is optimally 
		upper-bounded by $\sqrt{\frac{d}{2}}$ for even dimensions and 
		$\left(\frac{1}{d-1}+\frac{1}{d+1}\right)^{-\frac{1}{2}}$ for odd 
		dimensions \cite{GarciaPerez2006}. 
	\end{enumerate}
\end{enumerate}
The polar decomposition of the unital matrix $M$ generally 
splits it into two nonphysical constituents. Essentially, the unitary factor of $M$ ($R \in O(d^2-1)$ s.t. $R^{-1}M \geq 0$) can't generally be interpreted as a physically meaningful unitary operation. To see this, consider the following canonical Kraus decomposition:
\begin{align}
A_1    &= \left(\begin{array}{ccc}
\cos(\alpha) & 0 & 0 \\
0 & \cos(\alpha/2)e^{i\alpha^3/2}  & 0 \\
0 & 0 & \cos(\alpha/2) e^{-i\alpha^3/2}
\end{array} \right)~; \notag  \\
~A_2    &= \left(\begin{array}{ccc}
\sin(\alpha) & 0 & 0 \\
0 & -\sin(\alpha/2) e^{i (\alpha+\alpha^3/2)}  & 0 \\
0 & 0 & -\sin(\alpha/2) e^{-i (\alpha+\alpha^3/2)}
\end{array} \right).
\end{align}
The spectrum of the associated unital part $M$ is a subset of the spectrum of $A_1^* \otimes A_1 + A_2^* \otimes A_2 \in M_{d^2}(\mbb C)$ \footnote{$A_1^* \otimes A_1 + A_2^* \otimes A_2$ is the matrix acting on the column-vectorized density matrices, and has an extra eigenvalue of $1$ due the TP condition. Here the star $*$ denotes the complex conjugation, which is not to be confused with the star $\star$ used for the LK approximation.}. By 
expanding up to order $\alpha^4$, it is straightforward to show that the phase
factors of $M$ are all $\approx 1$ except for a single conjugate pair $\phi_{\pm}\approx {\rm exp}({\pm i 3 \alpha^3/2})$. This single pair can't be factored into any unitary process
since any non-trivial $V^*\otimes V$ contains at least two conjugate pairs. Hence, trying to cancel the rotating component of the spiraling action (see \cref{fig:spiral}) induced on $\vec{v}_{\pm}$ by $\phi_{\pm}$ would merely relocate the spiraling motion on an other pair of eigenvectors $\vec{v}'_{\pm}$ (or on multiple other pairs). 
To put it simply, spiraling is inherent to some
decoherent processes. To explicitly show this, we constructed an example
in which the rotation factors in the spirals
couldn't be accounted for by any physical unitary (without creating more spirals).

Separating a quantum channel $\mc A$ into a
composition of a physical unitary $\mc V$ and a decoherent operation $\mc D$ (i.e. $\mc A = 
\mc V \circ \mc D$ or $\mc A = \mc D \circ \mc V$) demands
a more careful surgery. If one were to 
allocate too many rotating components to the unitary factor,
$\mc V$ may fail to remain physical; 
on the other hand, allocating too little unitary action to $\mc V$
may leave the allegedly decoherent factor $\mc D$ with 
some physically reversible motion.
In fact, depending on the definition of decoherence, 
it is not even clear if such surgery is even possible. 
Here, we propose a definition of decoherence which 
allows to easily decompose any non-catastrophic quantum channel 
into a composition of a unitary channel with a decoherent 
one.

Consider a channel $\mc A$. Its LK operator $A_1 \in M_d(\mbb C)$ can be factored into a $d \times d$ unitary     
component $U$ multiplied with a positive semi-definite contraction $0<|A_1| \leq \mbb I_d$, i.e. $A_1 = V|A_1|$. This polar decomposition provides a geometric understanding
of the range of action of LK approximated channels on the space of 
quantum states. The absence of phase factors in the spectrum 
of $|A_1|$ motivates the following definition:

\begin{boxdefn}{decoherent channel}{decoherent}
	A non-catastrophic channel $\mc A$ is said to be decoherent if its LK operator is positive semi-definite:
	\begin{align}
	A_1 \geq 0~.
	\end{align}
\end{boxdefn}
From this definition immediately follows a unitary-decoherent decomposition
for quantum channels:

\begin{boxthm}{a polar decomposition for quantum channels}{decomp}
	Any non-catastrophic quantum channel $\mc A$ can be expressed as a composition of a unitary channel $\mc V$ with an decoherent channel $\mc D = \mc V^\dagger \circ \mc A $ (or $\mc D' = \mc A \circ \mc V^\dagger $):
	\begin{subequations}
		\begin{align}
		\mc A&=  \mc V \circ \mc D ~,\\
		\mc A&=  \mc D' \circ \mc V ~.
		\end{align}
	\end{subequations}
	In terms of LK approximation, we have:
	\begin{subequations}
		\begin{align}
		\mc A^\star(\rho) &=  A_1 \rho A_1^\dagger = V|A_1| \rho  |A_1|^\dagger V^\dagger~,\\
		\mc D^\star (\rho)&=  |A_1| \rho |A_1|^\dagger~,\\
		\mc D'^\star (\rho)&=  V |A_1| V^\dagger~ \rho~ V |A_1|^\dagger V^\dagger ~.
		\end{align}
	\end{subequations}
\end{boxthm}
\begin{proof}
	Under the composition $\mc V^\dagger \circ \mc A $, the 
	canonical Kraus operators $\{A_i\}$ of $\mc A$ are mapped
	to $\{ V^\dagger A_i\}$, since it preserves their orthonormality.
	Given the polar decomposition \mbox{$A_1= V|A_1|$}, it follows that the LK operator of $\mc V^\dagger \circ \mc A $ 
	is positive semi-definite.
\end{proof}

\subsection{The dynamics induced from decoherent channels as infinitesimal generators}
While the proof od \cref{thm:decomp} nearly trivially follows from 
\cref{defn:decoherent}, it remains to show that decoherent channels as we defined them deserve such an appellation. An interesting angle to initially justify our definition of decoherence
is to observe its contribution in the Gorini–-Kossakowski-–Sudarshan–-Lindblad (GKSL) equation \cite{Lindblad1976,GAS1976}.
Consider a time evolution dictated by instantaneous CPTP channels\footnote{This corresponds to the well-known Markovian regime.} with
(possibly time-dependent) canonical Kraus operators $\{A_k(t,dt)\}$:
\begin{align}
\rho(t+dt)= \sum_k A_k(t,dt) \rho(t) A_k^\dagger(t,dt)~.
\end{align}
Since $dt$ is infinitesimal, the instantaneous LK operator $A_1(t,dt)$
must be close to $\mbb I$, and can be expressed as
\begin{align}
A_1(t,dt)&= \exp \left(-iH(t) dt -P(t) dt\right) \notag \\
&= \mbb I - iH(t) dt- P(t) dt +O(dt^2)~, \label{eq:inst_LK}
\end{align}
where $H(t)$ is Hermitian and $P(t)$ is positive semi-definite. 
The TP condition can be expressed as
\begin{align}
\sum_k A^\dagger_k(t,dt)A_k(t,dt) = \mbb I ,
\end{align}
which combined with \cref{eq:inst_LK} yields
\begin{align}
P(t) dt = \frac{1}{2} \sum_{k \neq 1} A^\dagger_k(t,dt)A_k(t,dt) +O(dt^2).
\end{align}
This enforces the remaining instantaneous Kraus operators $A_{k \neq 1}(t,dt)$
to scale as $\sqrt{dt}$, and leaves us with 
\begin{align}\label{eq:Lind_equation}
\frac{d}{dt} \rho(t) = -i \left[H(t), \rho(t)\right] + \sum_{k\neq 1} L_k(t) \rho(t) L^\dagger_k(t)- \frac{1}{2}\left\{\sum_{k \neq 1} L_k^\dagger(t)L_k(t), \rho(t)\right\}~,
\end{align}
where 
\begin{align}
L_k(t):= \lim\limits_{dt \rightarrow 0} \frac{A_k(t,dt)}{\sqrt{dt}}~, \label{eq:Lind_limit}
\end{align}
and $[A,B]:= AB-BA$, $\{A,B\}:= AB+BA$ are respectively the well-known commutator and anticommutator. The fact that $\{A_k(t,dt)\}$ are canonical (hence orthogonal)
at every moment in time implies that $\inner{A_1(t,dt)}{A_{k \neq 1}(t,dt)}=0$, which
by using \cref{eq:inst_LK} results in
\begin{align}
\tr A_{k\neq 1}(t,dt) = -i dt \tr H(t) A_{k\neq 1}(t,dt) + dt \tr P(t) A_{k\neq 1}(t,dt)  +O(dt^2 \sqrt{dt})~.
\end{align}
This together with \cref{eq:Lind_limit} implies that 
\begin{align}
\tr L_k(t) = 0~.
\end{align}
Notice that the Lindblad operators featuring in a master equation generally do not have a 
zero trace, but since the master \cref{eq:Lind_equation} is derived from instantaneous 
\emph{canonical} Kraus operators, they do. That is, for every GKSL master
equation, there exists an alternate one, giving rise to the same dynamics, for which the Lindblad operators have a zero trace. This is an important feature for what follows. Let's re-express \cref{eq:Lind_equation} as a differential equation
acting on the column-vectorized states, ${\rm col} (\rho)$.

Using the property ${\rm col} (ABC)= C^T \otimes A~ {\rm col} (B)$, we have
\begin{align}\label{eq:Lind_master_col}
\frac{d}{dt} {\rm col} (\rho (t)) = \Bigg[\underbrace{-i\left(\mbb I \otimes H(t) - H^T(t) \otimes \mbb I \right)}_{i}  &\underbrace{-\frac{1}{2}\sum_{k \neq 1}\left(\mbb I \otimes L_k^\dagger(t)L_k(t) + (L_k^\dagger(t)L_k(t))^T \otimes \mbb I \right)}_{ii} \notag \\
 &+\underbrace{\sum_{k \neq 1} L^*_k(t) \otimes L_k(t)}_{iii}\Bigg] {\rm col} (\rho(t))~.
\end{align}
A quick calculation suffices to show that the three indicated terms are mutually orthogonal.
This means that their respective actions have no overlap. The first term should be familiar as it corresponds to the generator of unitary evolution. The remaining two terms are often referred to as the relaxation or decoherent part of the Lindbladian \cite{Ernst1987principles,Havel2003}. This integrates well with our notion of decoherence since the instantaneous channels are decoherent if and only if
the Hamiltonian is null at every moment in time:
\begin{align}
\exp \left(-i H(t) dt -P(t) dt \right) \geq 0 \Leftrightarrow H(t)=0~.
\end{align}
To formulate it otherwise, the Lindbladian consists solely of a decoherent part orthogonal to any commutator
if and only if the instantaneous channels are decoherent. An additional interesting remark
is that the LK approximation applied to the instantaneous channels essentially eliminates
the term $iii$, leaving only the commutator (term $i$) and the anticommutator (term $ii$).
In particular, the master equation with LK approximated instantaneous decoherent channels consists
of an anticommutator only:
\begin{align}
\frac{d}{dt} \rho(t)= - \{P(t), \rho(t)\}~.
\end{align}

When considered as infinitesimal perturbations from the identity, the channels that we refer to as ``decoherent'' correspond to the generators of the familiar class of decoherent master equations. 
While our notion of decoherence connects with previous physics literature in the infinitesimal case, it remains to show that our definition is also appropriate without taking such limit. 

\subsection{Further justifying our notion of decoherence}

Typically, quantum error channels are said to act decoherently 
if they exhibit a non-reversible deterioration. In turn, coherent
error channels correspond to a mishandling of information - 
which can in principle be reverted - rather than a loss of information. An additional expected property of decoherent operations is that they shouldn't
allow for coherent buildups such as in the case accumulating over-rotations.
Given $m$ non-catastrophic unitary channels $\mc V_i \approx \mc I$ with
\begin{align}
V_i =
\left(\begin{array}{cc}
\cos(\theta)  &-\sin(\theta) \\
\sin(\theta) &  \cos(\theta) 
\end{array} \right) ~,
\end{align}
the infidelity grows faster than linearly  (let the composition $\mc V_{m:1}$ be non-catastrophic so that $ m\theta \leq \pi/4$) \cite{Dugas2016}:
\begin{align}
1-\sqrt{\Phi(\mc V_{m:1},\mc I)} = 
1-\cos\left(m \theta_i\right) 
\geq m \left(1-\cos(\theta)\right) = \sum_i \left(1-\sqrt{\Phi(\mc V_i,\mc I)}\right)~.
\end{align}

As an intuitive pair of properties of our decoherent channels, we show that
\begin{enumerate}[i.]
	\item The average process fidelity of decoherent error channels cannot 
	be substantially recovered by any unitary (quasi-monotonicity).
	\item The evolution of the infidelity of a circuit composed of
	decoherent operations is (approximately) at most additive in the individual infidelities. There is no substantial coherent buildup.
\end{enumerate} 
\begin{boxthm}{two features of decoherence}{fid_evo_1}
	Consider $m$ non-catastrophic decoherent channels $\mc D_i$ and any non-catastrophic unitary channel $\mc V$. Then,
	\begin{subequations}
		\begin{align}
		\Phi(\mc V \circ \mc D_{m:1}, \mc I) &\leq  \min_i \Phi(\mc D_{i}, \mc I)\notag\\
		&~+\frac{1}{2} \left(\sum\limits_{i=1}^m (1-\Upsilon(\mc D_i^\star))\right)^2+
		(1- \Phi(\mc V \circ \mc D_{m:1}^\star, \mc I)) 		\sum \limits_{\substack{i=1} }^{m}\left(1- \Upsilon(\mc D_i^\star)\right) \tag{Quasi-monotonicity}\\
		1-\Phi(\mc V \circ\mc D_{m:1}, \mc I)&\leq \left(1- {\Phi(\mc V, \mc I)}\right)+ \sum\limits_{i=1}^m (1-\Phi(\mc D_i, \mc I)) \notag \\
		&~~+ (1- \Phi(\mc V, \mc I))^2+ \sum\limits_{i=1}^m (1- \Phi(\mc D_i^\star, \mc I))^2 \notag \\
		&~~+\sum\limits_{i=1}^m (1- \Phi(\mc D_i, \mc I))(1-\Upsilon^2(\mc D_i))
		\tag{Quasi-subadditivity property}~
		\end{align}
	\end{subequations}
\end{boxthm}



\section{Behavioral signatures of coherence and decoherence}\label{sec:assesment}
The introduction in the previous section of the dichotomy between coherence and decoherence, together with the demonstration of a polar decomposition for quantum channels wasn't void of ulterior motives. In this section, we leverage the intrinsic
differences between coherent and decoherent channels to explore the behavior of the average process fidelity and the unitarity as 
circuits grow in depth. 
Before we begin such investigation, however, 
let's first make a side step 
to define various classes of operations which will 
harmonize with our notion of decoherence. 

\subsection{Extremal dephasers, extremal unitaries, and equable error channels}\label{sec:dephasers}
The non-catastrophic condition still leaves room for pathological noise scenarios. 
We highlight two extreme (unrealistic) types of channel; the first is of decoherent nature, 
and the second is purely unitary.

\subsubsection{Extremal dephasers}
For a channel $\mc A$ to be non-catastrophic, the singular values
of its LK operator $\sigma_i(A_1)$ must nearly average to $1$, but nothing else 
constrains their distribution. Consider a $10$-qubit
error $\mc A$ that essentially acts as identity on all operators in $M_d(\mbb C)$, 
but cancels any phase between $|0\rangle$ and $| i 
\rangle$ for $i \neq 0$ (that is,  $| 0 \rangle \langle i |, | i  \rangle 
\langle 0 | \rightarrow 0 $ for $i \neq 0$). It is easily shown that the LK 
operator is $A_1 = \sum_{i\neq 0} |i \rangle \langle i|$; this is an instance of what 
we call an ``extremal dephaser''. An extremal dephaser is defined as a channel
for which there exists a singular value $\sigma_j \in \{\sigma_i(A_1)\}$ (in our example, it is $\sigma_0=0$) that 
deviates from $1$ by much more than the average perturbation:
\begin{align}\label{eq:extremal_dephaser}
1- \sigma_j \gg 1- \mbb E_i[\sigma_i]~.
\end{align}
To obey \cref{eq:extremal_dephaser}, channels
must involve excessively strong\footnote{Relative to other decoherent mechanisms.} dephasing mechanisms between a small 
number of states and the rest of the system\footnote{This is entirely different 
	than: ``excessively strong dephasing mechanisms between a \emph{small 
		subsystem} and the rest of the system'', which we already discarded 
	through the non-catastrophic assumption.}. 
Let's come back to our example:
a quick calculation shows that $\mc A$ has an infidelity of around $O(2^{-10}) = O(10^{-3})$:
extremal dephasers can have a high average fidelity; they are not ruled out by
the non-catastrophic assumption. However, based on realistic grounds, one might discard such scenarios by assuming that the perturbations of the singular values $|1- \sigma_j|$ remain comparable to the average perturbation $\mbb E[1-\sigma_i(A_1)]$. 
Indeed, most physically motivated noise mechanisms -- such as unitary, amplitude 
damping and stochastic channels\footnote{A stochastic channel has (up to 
	constant factors) unitary operations as canonical Kraus operators and has a 
	LK 
	operator proportional to the identity. Examples of orthogonal unitary bases 
	include the Heisenberg-Weyl operators, and the n-fold tensor product of 
	Paulis. 
	Standard dephasing channels are a special case of stochastic channels were 
	the 
	unitaries are simultaneously diagonalizable (i.e. they all commute).} -- 
perturb 
the singular values of $A_1$
in a rather homogeneous way (see \cref{tab:channels}). 

\begin{figure}[!b]
	\centering
	{\caption{Singular values $\sigma_i$ -- plotted as purple circles -- of the 
	$10^3 \times 10^3$ LK operator $A_1$ of an extremal dephaser $\mc A$. The dashed line corresponds to the average $\mbb E_i[\sigma_i]= 0.9989(1)$. The green shaded region covers a standard deviation $\text{SD}[\sigma_i]=0.0032(1)$ below the average. In this example, the standard deviation is roughly three times greater that the average deviation $1-\mbb E_i[\sigma_i]= 0.0011(1)$; that is, the WSE decoherence constant (see \cref{defn:wse}) is $\gamma_{\rm decoh}=3.0(1)$, which is an order of magnitude smaller than $1 / \sqrt{\mbb E[1-\sigma_i]} = 30.7(1)$. From \cref{eq:gamma_D_cond}, $\mc A$ is equable in the wide-sense.
	 There are five singular values situated around $0.955$,  meaning that $1- \sigma_j$ can be more than forty times larger that the average deviation (i.e. $\Gamma_{\rm decoh} = 41(1)$). While these extreme deviations are excluded by the equability condition, their small impact on the standard deviation allows $\mc A$ to be WSE. 
	}\label{fig:wse}}
	{\includegraphics[width=.9\textwidth]{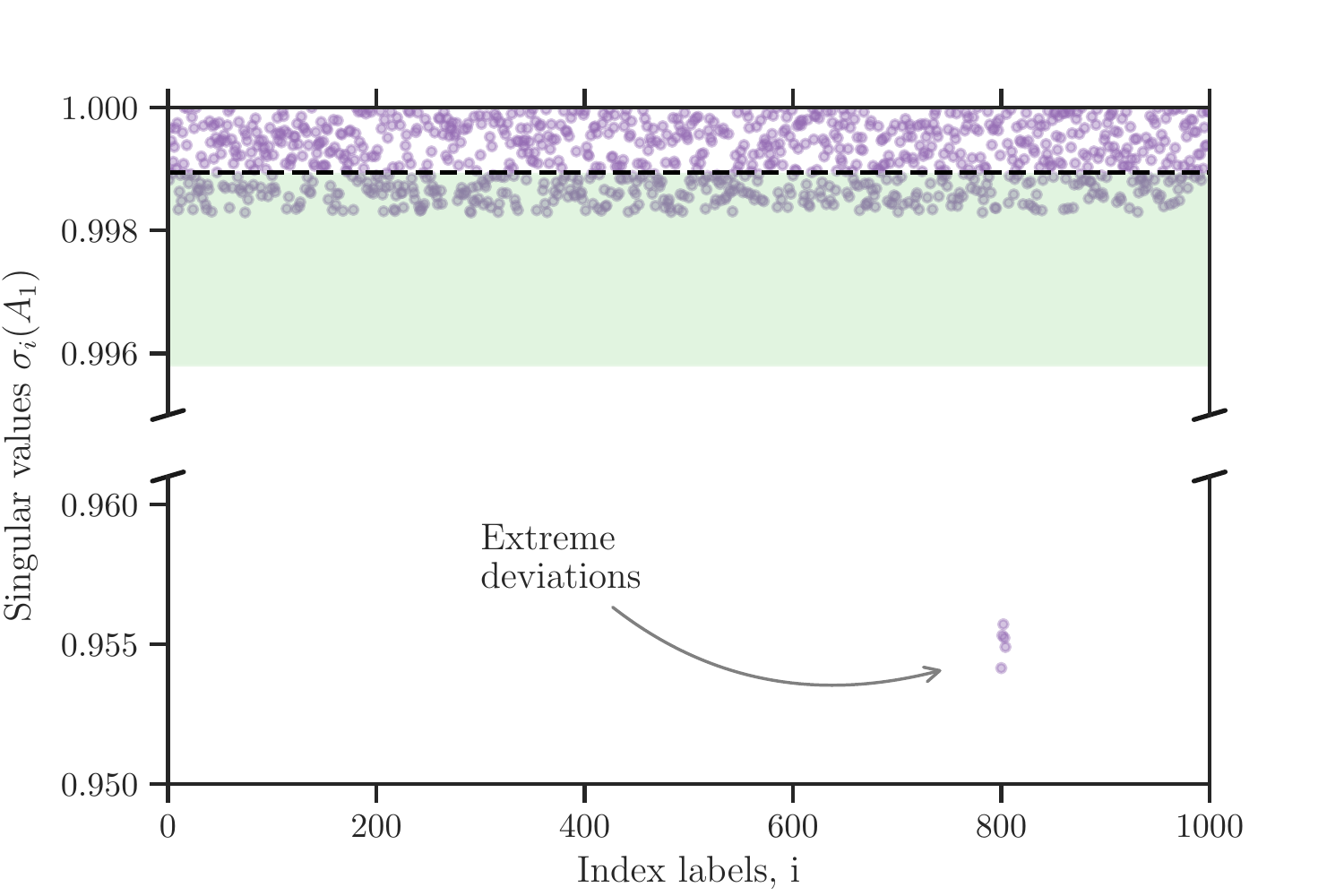}}
\end{figure}

\subsubsection{Extremal unitaries}
The same argument that was made about the singular values of $A_1= V|A_1|$
, which are the eigenvalues of its positive semidefinite factor $|A_1|$,
can be made for the eigenvalues of the unitary factor $V$. To mimic our
previous example, consider a $10$-qubit
unitary error $\mc V$ that essentially acts as identity on operators in $M_d(\mbb C)$, but
maps $| 0 \rangle \langle i |\rightarrow -| 0 \rangle \langle i |$ 
$| i \rangle \langle 0 |\rightarrow -| i \rangle \langle 0 |$ for $i \neq 0$. It is easily shown that the LK 
operator is $V =-|0 \rangle \langle 0|+ \sum_{i\neq 0} |i \rangle \langle i|$; this is an instance of what 
we call an ``extremal unitary''. 
An extremal unitary is defined as a unitary error $\mc V$
for which there exists an eigenvalue $\lambda_j \in \{\lambda_i(V)\}$ (in our example, it is $\lambda_0=-1$) that 
deviates from $1$ by much more than the average perturbation. 
An easy way to make this precise is to fix the phase of $V$ such that 
$\tr V \in \mbb R_+$, and project the eigenvalues on the real axis (this is easy to picture on an Argand diagram):
\begin{align}\label{eq:extremal_unitary}
	1- \text{Re}\{\lambda_j\} \gg 1- \mbb E_i[\text{Re}\{\lambda_i\}] = 1-\tr V /d.
\end{align}
To obey \cref{eq:extremal_unitary}, the unitary error must result
from a strong alteration made to the targeted Hamiltonian. Indeed, as
a simple Taylor expansion can confirm, small 
perturbations from the intended Hamiltonian cannot yield
an extremal unitary error. Just as for extremal dephasers, 
extremal unitaries can have a high average fidelity, yet can be reasonably discarded.
The perturbations $1-\text{Re}\{\lambda_i\}$ are expected to be comparable to
the average perturbation $1- \mbb E_i[\text{Re}\{\lambda_i\}]$ (here, $\tr V \in \mbb R_+$).

\subsubsection{Equable error channels}
In this paper, we qualify 
as ``equable'' the non-catastrophic error channels $\mc A = \mc V \circ \mc D $
for which the factors $\mc D$ and $\mc V$ are not extremal. 
Notice that the equability assumption ensures a unique polar decomposition since the LK operator is 
guaranteed to be full rank.

While ruling out extremal error channels seems reasonable, we also define a weaker condition based
on the variance of the perturbations. 
\begin{boxdefn}{\textbf{E}quable error channels}{wse}
	Consider a non-catastrophic error channel $\mc A =\mc V \circ \mc D $ with
	LK operator $A_1 = V |A_1| $.
	Let $\{\sigma_i\}$  be the singular values of $A_1$ and $\{\lambda_i\}$ be the eigenvalues 
	of $V$ for which the phase is fixed such that $\tr V \in \mbb R_+$. 
	We define the strict-sense equability (SSE) decoherence and coherence constants $\Gamma_{\rm decoh}, \Gamma_{\rm coh}$ as:
	\begin{subequations}
		\begin{align}
		1 - \min_j \sigma_j &= \Gamma_{\rm decoh} \mbb E[1-\sigma_i]~, 
		\label{eq:sse_decoh_cst}	
		\\
	   	1 - \min_j  \text{Re}\{\lambda_j\} &= \Gamma_{\rm coh} \mbb E[1- \text{Re}\{\lambda_i\}] ~.
		\label{eq:sse_coh_cst}
		\end{align}
	\end{subequations}
	A non-catastrophic error channel is said to be equable (in the strict sense) if 
	\begin{align}\label{eq:Gamma_D_cond}
	\Gamma_{\rm decoh} &\ll  1 / \sqrt{\mbb E[1-\sigma_i]}~, \\
	\Gamma_{\rm coh} &\ll 1 / \sqrt{\mbb E[1- \text{Re}\{\lambda_i\}]}~.
	\label{eq:Gamma_U_cond}
	\end{align}
	Analogously, we define the wide-sense equability (WSE) decoherence and coherence constants $\gamma_{\rm decoh}, \gamma_{\rm coh}$ as:
	\begin{subequations}
			\begin{align}
			{\rm SD}[\sigma_i] &= \gamma_{\rm decoh} \mbb E[1-\sigma_i]~, 
			\label{eq:wse_decoh_cst}	
			\\
			{\rm SD}[\text{Re}\{\lambda_i\}]&= \gamma_{\rm coh} \mbb E[1- \text{Re}\{\lambda_i\}] ~,
			\label{eq:wse_coh_cst}
			\end{align}
	\end{subequations}
	where $\rm SD$ denotes the standard deviation. A non-catastrophic error channel is said to be equable in the wide sense if 
	\begin{align}\label{eq:gamma_D_cond}
		\gamma_{\rm decoh} &\ll  1 / \sqrt{\mbb E[1-\sigma_i]}~, \\
		\gamma_{\rm coh} &\ll 1 / \sqrt{\mbb E[1- \text{Re}\{\lambda_i\}]}~.
		\label{eq:gamma_U_cond}
	\end{align}
\end{boxdefn}
First notice that ruling out extremal errors is directly imposed by the equability condition (in the strict sense). Obviously, equability implies wide-sense equability, since by construction
\begin{subequations}
	\begin{align}
		\gamma_{\rm coh} &\leq \Gamma_{\rm coh}~, \\
		\gamma_{\rm decoh}& \leq \Gamma_{\rm decoh}~.
	\end{align}
\end{subequations}
Of course, the converse doesn't hold (see \cref{fig:wse} for an example), 
although such pathological cases must involve extremal channels.
The motivation behind the weaker definition is not physical, but mathematical.
The results exhibited in \cref{thm:uni_evo_2,thm:fid_evo_2,thm:max_cor,thm:fid_evo_3,thm:max_cor_multi} solely rely on 
the WSE constants rather than on the realistically slightly larger SSE constants.

\begin{table}[!h]
	\centering
	\begin{tabular}{|l| C{10em} |C{6em}|C{8em}|}
		\hline
		Error channel & Type of error & LK operator & Coherence level, $r_{\rm coh}/r$\\
		\hline 
		\hline 
		Depolarizing & Decoherent, SSE & $A_1 \propto \mbb I$ & $O(r)$ \\
		\hline 
		Standard dephasing & Decoherent, SSE & $A_1 \propto\mbb I$& $O(r)$ \\
		\hline 
		Stochastic & Decoherent, SSE& $A_1 \propto \mbb I$ & $O(r)$ \\
		\hline
		Amplitude damping & Decoherent, realistically SSE & $A_1 \geq 0$ & $O(r)$ \\
		\hline
		Unitary & Coherent, realistically SSE  & $A_1 = V$& 1 \\
		\hline
		General SSE & Contains a coherent and decoherent factor & $A_1 = V|A_1|$ & $\frac{d^2-|\tr V|^2}{d^2-|\tr A_1|^2}+O(r)$ \\
		\hline
	\end{tabular} 
	\caption{Categorization of different well-known error channels. Many canonical error mechanisms fall under the ``decoherent'' appellation, except for unitary errors, of course. The coherence level is negligible for decoherent channels, and $1$ for coherent errors. In the intermediate regime, the coherence level can vary between $0$ and $1$. It only makes sense to discuss about the coherence level when errors are equable (at least in the wide sense). \label{tab:channels}}
\end{table}

\subsection{Reasoning about $\Upsilon$}
Now that we have defined (wide-sense) equable errors, we are ready to 
express a first decay law:
\begin{boxthm}{unitarity's decay law}{uni_evo_2}
	Consider $m$ non-catastrophic channels $\mc A_i$. Then $\Upsilon(\mc A_{m:1})$
	has the following properties:
	\begin{align}
	\Upsilon(\mc A_{m:1}) &\leq \min_i \Upsilon(\mc A_i)+ (1-\Upsilon^2(\mc A_{m:1}))^2/\sqrt{2} \tag{Quasi-monotonicity} \\
	1-\Upsilon(\mc A_{m:1})  &\leq 
	\sum \limits_{i=1}^m \left(1-\Upsilon(\mc A_i)\right)+ (1-\Upsilon^2(\mc A_i))^2 \tag{Quasi-subadditivity property}
	\end{align}
	The quasi-monotonicity is almost saturated by extremal channels. 
	If we introduce the WSE decoherence constants $\gamma_{\rm decoh}(\mc A_i) \leq \gamma_{\rm decoh}$, we obtain:
	\begin{align}
\abs{\Upsilon(\mc A_{m:1})- \prod_i \Upsilon(\mc {A}_i)} &\leq 
(1- \Upsilon(\mc A_{m:1}^\star))^2 +  \sum_{j=1}^m  (1- \Upsilon(\mc {A}^\star_j))^2 \notag \\
&+\gamma_{\rm decoh}^2 \sum_{i=1}^{m} \left(1-\sqrt{\Upsilon(\mc {A}_i^\star)}\right)^2 \notag \\
&+2 \gamma_{\rm decoh}^2 \left(\sum_{i=1}^m \left(1-\sqrt{\Upsilon(\mc {A}_i^\star)}\right)\right)^2+{\text{H.O.T.}}~
\end{align}
	If the channels are equable, $\Upsilon(\mc A_{m:1})$ is essentially a multiplicative decay.
\end{boxthm}
Of course, those results can be immediately translated in terms of unitarity
by using \cref{eq:chi_2_uni}.
Without using the LK approximation, showing the monotonicity of the unitarity
can be difficult, since quantum channels aren't contractive maps; 
going to the LK picture fixes this issue since Kraus operators 
are contractions. 
Quasi-multiplicativity is
another way of stating that the unitarity of a composition
essentially behaves as a multiplicative decay involving the unitarity of 
individual components:
\begin{align}
u(\mc A_{m:1}) \approx \frac{d^2 \prod_{i=1}^{m}\Upsilon^2(\mc 
	A_i)-1}{d^2-1}~.\label{eq:uni_decay}
\end{align}
\Cref{eq:uni_decay} should be seen as a staple of wide-sense equability; deviations from this 
behavior indicates the presence of extremal dephasers. 

The quasi-multiplicativity of $\Upsilon$ is not the only decay law that occurs 
in the equable scenario. Recall that to motivate our definition of decoherence, we 
initially showed the quasi-monotonicity and quasi-subadditivity property of the 
process fidelity of decoherent compositions (\cref{thm:fid_evo_1}).
By introducing the equability condition we get a stronger assertion:
\begin{boxthm}{fidelity's decay law (for decoherent compositions)}{fid_evo_2}
	Consider $m$ non-catastrophic, decoherent channels $\mc D_i$ (with target $\mc I$) with WSE decoherence constants $\gamma_{\rm decoh}(\mc D_i) \leq \gamma_{\rm decoh}$. 
	Then, $\Phi(\mc D_{m:1}, \mc I)$ is bounded as follows:
	\begin{align}
	&\abs{\Phi(\mc D_{m:1}, \mc I) - \prod_{i=1}^{m}  \Phi(\mc D_i,\mc I)} \leq \Bigg[
	\frac{1}{2}\left(\sum\limits_{i=1}^m (1-\Upsilon(\mc D_i^\star))\right)^2 \notag \\
	&~~+(1- \Phi(\mc D_{m:1}^\star, \mc I))\sum \limits_{\substack{i=1} }^{m}\left(1- \Upsilon(\mc D_i^\star)\right)
	~~+\sum\limits_{i=1}^m  \left(  1-\Upsilon(\mc D_i^\star) \right)\left(  1-\Phi(\mc D_i, \mc I) \right)
	\notag \\
	&+\gamma_{\rm decoh}^2 \prod_{i=1}^{m} \sqrt{\Phi(\mc D_i^\star,\mc I)} \left(\sum\limits_{i=1}^{m} \left(1-\sqrt{\Phi(\mc D_i^\star, \mc I)}\right) \right)^2 \Bigg] +{\rm H.O.T.}
	\label{eq:fid_evo_equ_dec}
	\end{align}
	If the channels are WSE, $\Phi(\mc D_{m:1}, \mc I)$ is essentially a multiplicative decay.
\end{boxthm}
Using the simple relation between $F$ and $\Phi$ (\cref{eq:fid}) we come to this observation:
the average gate fidelity of a composition of non-catastrophic decoherent equable channels behaves almost exactly as a multiplicative decay in the average process fidelity of individual components, that is
\begin{align}
F(\mc D_{m:1}, \mc I) \approx \frac{d \prod_{i=1}^{m} \Phi(\mc D_i, \mc I)+1}{d+1}~.
\end{align}
The decay becomes exact with the depolarizing channel 
$\mc P_p(\rho) = p \rho+ (1-p)(\tr \rho) \mbb I_d /d$, 
which is a celebrated example of a decoherent
operation. 

The two decay laws expressed in \cref{thm:uni_evo_2,thm:fid_evo_2} are in fact 
describing the same observation. Let $\mc A$ have an equable error and a polar decomposition $\mc V \circ \mc D$.
As shown in the following theorem, $\Upsilon(\mc A)$ can be interpreted as 
the maximal process fidelity of $\mc A$ to the target $\mc U$ under unitary corrections, 
or equivalently as the process fidelity of the decoherent factor $\mc D$ to the identity:
\begin{align}
	\Upsilon(\mc A) \approx \Phi(\mc D, \mc I) \approx \max \limits_{W \in SU(d)} \Phi(\mc W \circ \mc A, \mc U)~.
\end{align}
\begin{boxthm}{$\Upsilon$ as the process fidelity of the decoherent factor}{max_cor}
	Consider a non-catastrophic channel $\mc A = \mc V \circ \mc D $ with unitary target $\mc U$. Then, the maximal unitary correction of $\mc A$ (in terms of $\Phi$) is approximately bounded by the interval $\left[\Upsilon^2(\mc A), {\Upsilon(\mc A)}\right]$:
	\begin{subequations}
		\begin{align}
		\max \limits_{W \in SU(d)} \Phi(\mc W \circ \mc A, \mc U) &\leq {\Upsilon(\mc A)} + \frac{3}{2}(1-\Upsilon^2(\mc A))^2 ~, \label{eq:upper_phi_corr} \\
		\max \limits_{W \in SU(d)} \Phi(\mc W \circ \mc A, \mc U) &\geq \Upsilon^2(\mc A) - (1-\Upsilon^2(\mc A))^2~. \label{eq:lower_phi_corr}
		\end{align}
	\end{subequations}
		Moreover, if we introduce the WSE decoherence constant $\gamma_{\rm decoh}$, we obtain:
	\begin{align}
	\max \limits_{W \in SU(d)} \Phi(\mc W \circ \mc A, \mc U) \geq {\Upsilon(\mc A)} - (1+\gamma_{\rm decoh}^2) \left(1-{\Upsilon^2(\mc A)}\right)^2~.
	\end{align}
	A quasi-maximal choice of unitary correction consists in $\mc W = \mc U \circ \mc V^\dagger$.
\end{boxthm}
In terms of other figures of merit, wide-sense equability ensures a quasi-one-to-one correspondence 
between the maximal average gate fidelity (through a unitary correction) and the unitarity through:
\begin{align}
\max \limits_{W \in SU(d)} F(\mc W \circ \mc A, \mc U) \approx F( \mc D, \mc I) \approx \frac{\sqrt{(d^2-1)u(\mc A)+1}+1}{d+1}~.
\end{align}

\subsection{The coherence level}
Let's extend \cref{thm:fid_evo_2} by appending a coherent operation to the decoherent composition:
\begin{boxthm}{the average process fidelity of equable compositions}{fid_evo_3}
	Consider $m$ non-catastrophic, decoherent error channels $\mc D_i$ (with target $\mc I$) with WSE decoherence
	constants $\gamma_{\rm decoh}(\mc D_i) \leq \gamma_{\rm decoh}$.
	Moreover, consider
	a non-catastrophic unitary error channel $\mc V$ with WSE coherence
	constant $\gamma_{\rm coh}$. Then, $\Phi(\mc V \circ \mc D_{m:1}, \mc I)$ is bounded as follows:
	\begin{align}
	&\abs{\Phi(\mc V \circ \mc D_{m:1}, \mc I) - \Phi(\mc V, \mc I) \prod_{i=1}^{m} \Phi(\mc D_i, \mc I)}   \leq  \Bigg[
	 \frac{1}{2}\left(\sum\limits_{i=1}^m (1-\Upsilon(\mc D_i^\star))\right)^2
	\notag\\
	&+(1- \Phi(\mc V \circ \mc D_{m:1}^\star, \mc I))\sum \limits_{\substack{i=1} }^{m}\left(1- \Upsilon(\mc D_i^\star)\right)+\sum\limits_{i=1}^m \left(  1-\Upsilon(\mc D_i^\star) \right)\left(  1-\Phi(\mc D_i, \mc I) \right)
	\notag \\
	&+ 2{\gamma_{\rm decoh}}{\gamma_{\rm coh}}\left(1-\sqrt{\Phi(\mc V, \mc I)}\right)\sum\limits_{i=1}^{m}	\left(1-\sqrt{\Phi(\mc{D}_i^\star, \mc I)}\right) \notag \\
	&+{\gamma_{\rm decoh}^2}\left(\sum\limits_{i=1}^{m} \left(1-\sqrt{\Phi(\mc D_i^\star, \mc I)}\right) \right)^2 \Bigg]+{\rm H.O.T.}
	\label{eq:quasi_mult_U}
	\end{align}
	If the errors are WSE, then $\Phi(\mc V \circ \mc D_{m:1}, \mc I)$ is essentially multiplicative.
\end{boxthm}
Let's unfold this result one step at a time. First, consider 
\cref{eq:quasi_mult_U} for $m=1$. Let $\mc A$ be a channel with target $\mc U$
and polar decomposition $\mc V \circ \mc D$. $\mc W:=\mc U^{-1}\circ\mc V$ is a unitary error. Hence, it follows from \cref{thm:fid_evo_3,thm:max_cor} that
\begin{align}
\Phi(\mc A, \mc U)=	\Phi(\mc W \circ \mc D, \mc I)\overset{\rm thm. 8}{\approx} \Phi(\mc W, \mc I) \Phi(\mc D, \mc I) = \Phi(\mc V, \mc U) \Phi(\mc D, \mc I) \overset{\rm thm. 7}{\approx}  \Phi(\mc V, \mc U) {\Upsilon(\mc A)}~. \label{eq:mult}
\end{align}
There are two factors that compound to the average process fidelity: $\Phi(\mc V,\mc U)$ relates to a coherent contribution to the total infidelity, while $\Phi(\mc D, \mc I) \approx {\Upsilon(\mc A)}$ depicts a decoherent one.
For those who are more familiar with the infidelity $r(\mc A, \mc U)$, \cref{eq:mult} can be reformulated as\footnote{The transition from \cref{eq:mult} to \cref{eq:split_inf} simply involves using the approximation $(1-\delta_1)(1-\delta_2) \approx 1-\delta_1-\delta_2$ for small $\delta_i$.} (up to $O(r^2)$):
\begin{align}\label{eq:split_inf}
r(\mc A, \mc U) \approx \underbrace{r(\mc V, \mc U)}_{\text{Coherent infidelity}} + \underbrace{r(\mc D, \mc I)}_{\text{Decoherent infidelity}} = r_{\rm coh}+ r_{\rm decoh}~.
\end{align}
The channel average infidelity of a channel can be split into a sum of a coherent and decoherent terms (given equable errors). $r_{\rm decoh}$ is not substantially correctable through any composition, and can be obtained from the unitarity alone:
\begin{align}\label{eq:r_decoh_uni}
r_{\rm decoh} = \frac{d-\sqrt{(d^2-1)u(\mc A)+1}}{d+1} + O(r^2) =\frac{d}{d+1}\left( 1-\Upsilon(\mc A) \right)+O(r^2)~.
\end{align} 
$r_{\rm coh}$ can be corrected through a composition with a unitary (see \cref{thm:max_cor}). 
\Cref{{eq:split_inf}} motivates the definition of \emph{coherence level} as the fraction of
the infidelity that is associated to coherence. It can be obtained by combining the infidelity and the unitarity through:
\begin{align}
	\frac{r_{\rm coh}}{r} = 1-\frac{d-\sqrt{(d^2-1)u(\mc A)+1}}{(d+1)r(\mc A, \mc U)} +O(r) = \frac{1-\Upsilon(\mc A)}{1-\Phi(\mc A, \mc U)}+O(r)
\end{align}
 Similarly, the decoherence level is defined as $r_{\rm decoh}/r$. 
\Cref{eq:split_inf} strengthens the insight behind the notion of coherence level introduced (under different appellations)
in \cite{Feng2016,Yang2019}. In those previous works, the RHS of
\cref{eq:split_inf} is generally depicted as a lower bound on the infidelity, which can be reduced to $r_{\rm decoh}$ through a unitary correction. The (approximate) equality -- which is much more valuable since it provides an upper bound on $r$ -- is shown for single qubit case in \cite{Feng2016} using the polar decomposition of the action on Bloch sphere. 
Here, we have shown the (approximate) equality (in the equable scenario) for all dimensions using the polar decomposition of LK operators.

\subsection{Bounding the worst and best case fidelity of a circuit}
Now, let's revisit \cref{thm:fid_evo_3} for general circuit depth $m$. This will allow us to identify the worst and best case fidelity of a circuit.
Consider $m$ channels $\mc A_i$ with target $\mc U_i$ and polar decomposition
$\mc D_i \circ \mc V_i$. The circuit $\mc A_{m:1}$ can be re-expressed as
\begin{align}
\mc A_{m:1}&= \mc V_{m:1} \circ (\mc V_{m:1})^\dagger \circ \mc D_m \circ \mc V_{m:1} \circ \cdots \circ (\mc V_{2:1})\circ \mc V_1^\dagger\circ \mc D_1 \circ \mc V_1 
= \mc V_{m:1} \circ \mc D'_{m:1}~,
\end{align}
where $\mc D_k':= (\mc V_{k:1})^\dagger \circ \mc D_k \circ \mc V_{k:1}$ are decoherent channels with the same fidelity as $\mc D_k$. This means that:
\begin{align}
\Phi(\mc A_{m:1}, \mc U_{m:1}) \overset{\rm thm. 8}{\approx} \Phi(\mc V_{m:1}, \mc U_{m:1}) \prod_{i=1}^m \Phi(\mc D_i, \mc I) \overset{\rm thm. 7}{\approx} \Phi(\mc V_{m:1}, \mc U_{m:1}) \prod_{i=1}^m {\Upsilon(\mc A_i)} ~. \label{eq:phi_evo}
\end{align}
In this last expression, we clearly see that the evolution of $\Phi$ is factored into a decoherent decay multiplied by a function $\Phi(\mc V_{m:1}, \mc U_{m:1})$ which captures the fidelity of a purely coherent process. This is already an interesting realization: since the decoherent decay is fixed, all the freedom in the evolution of the fidelity is contained in the coherent factors. 
An assessment concerning 
the circuit's average process fidelity must rely 
on a characterization of coherent effects. Since we know that such effects 
are correctable through composition, we first get: 
\begin{boxthm}{maximal average process fidelity of channel compositions}{max_cor_multi}
	Consider $m$ non-catastrophic channels $\mc A_i$ with respective unitary targets $\mc U_i$ and polar decompositions $\mc A_i = \mc V_i \circ \mc D_i $. Let the WSE decoherence constants be $\gamma_{\rm decoh}(\mc D_i) \leq \gamma_{\rm decoh}$.
	Then, the maximal unitary correction of the composition $\mc A_{m:1}$ is bounded as follows:
	\begin{subequations}
		\begin{align}
		&\scalemath{0.9}{\max \limits_{W \in SU(d)} \Phi(\mc W \circ \mc A_{m:1}, \mc U_{m:1}) - \prod_{i=1}^{m} {\Upsilon(\mc A_i)} \leq \Bigg[ \frac{1}{2} \left(\sum\limits_{i=1}^m (1-\Upsilon(\mc A_i^\star))\right)^2
		+ \sum_{i=1}^{m} (1-\Upsilon(\mc {A}^\star_{i}))^2}\notag \\
		&~~~\scalemath{0.9}{+\left(\sum \limits_{\substack{i=1} }^{m}\left(1- \Upsilon(\mc A_i^\star)\right)\right)\left(1- \prod_{i=1}^{m} {\Upsilon(\mc {A}_{i})}\right) + 2 \gamma_{\rm decoh}^2 \left(\sum_{i=1}^m \left(1-\sqrt{\Phi(\mc {D}_i^\star,\mc I)}\right)\right)^2 \Bigg] +{\rm H.O.T.}}\\		
		&\scalemath{0.9}{\max \limits_{W \in SU(d)} \Phi(\mc W \circ \mc A_{m:1}, \mc U_{m:1}) - \prod_{i=1}^{m} {\Upsilon(\mc A_i)} \geq
		\Bigg[- \gamma_{\rm decoh}^2 \sum_{i=1}^{m} \left(1-\sqrt{\Phi(\mc {D}_i^\star,\mc I)}\right)^2
		-\sum_{i=1}^{m} (1-\Upsilon(\mc {A}^\star_{i}))^2} \notag \\
		&~~~\scalemath{0.9}{- \gamma_{\rm decoh}^2 \prod_{i=1}^{m} \sqrt{\Phi(\mc D_i^\star,\mc I)} \left(\sum\limits_{i=1}^{m} \left(1-\sqrt{\Phi(\mc D_i^\star, \mc I)}\right) \right)^2 \Bigg] +{\rm H.O.T.}}
		\end{align}
	\end{subequations}
	For equable errors, the maximal unitary correction of the composition $\mc A_{m:1}$ is essentially $\prod_{i=1}^{m} {\Upsilon(\mc A_i)}$.
	A quasi-optimal choice of unitary correction is $\mc W = \mc U_{m:1} \circ (\mc V_{m:1})^\dagger$.
\end{boxthm}
In short, the average gate fidelity of a composite circuit is upper bounded by a decaying envelope which is closely
prescribed by the decoherent factors of its individual components:
\begin{align}
\max \limits_{W \in SU(d)}F(\mc W \circ \mc A_{m:1}, \mc U_{m:1}) \approx\frac{d \prod_{i=1}^m \Phi(\mc D_i,\mc I)+1}{d+1} \approx \frac{d \prod_{i=1}^m {\Upsilon(\mc A_i)}+1}{d+1}~.
\end{align}
This unforgiving behavior harmonizes well with the more typical comprehension
of decoherence as a limiting process.

To find the worst possible $\Phi(\mc A_{m:1}, \mc U_{m:1})$, it suffices
to use a lower bound for the coherent factor $\Phi(\mc V_{m:1}, \mc U_{m:1})$. 
This is partially done in \cite{Dugas2016}, where the inequality
\begin{align}
\Phi(\mc V_{m:1}, \mc U_{m:1}) \geq  \cos^2\left(\sum\limits_{i=1}^m \arccos\left(\sqrt{\Phi(\mc V_{i}, \mc U_{i})}\right)\right)~
\end{align}
is shown to be saturated in even dimensions. For odd dimensions, we find the following saturated bound:
\begin{align}
\Phi(\mc V_{m:1}, \mc U_{m:1}) \geq  \left(\frac{(d-1)\cos\left(\sum_{i=1}^{m} \arccos\left(\frac{d \sqrt{\Phi(\mc V_i, \mc U_i)}-1}{d-1}\right)\right) +1}{d}\right)^2~.
\end{align}
\begin{proof}
	The generalization to odd dimensions almost immediately follows by looking at the saturation case in even dimensions, which consists of commuting unitary errors of the form
	\begin{align}
	\left(\begin{array}{cc}
	\cos(\theta_i)  &-\sin(\theta_i) \\
	\sin(\theta_i) &  \cos(\theta_i) 
	\end{array} \right)\otimes \mbb I_{d/2}~.
	\end{align}
	In the odd dimension case, it suffices to always pick the global phase to fix the first eigenvalue of $V_{m:1} (U_{m:1})^{-1}$ to $1$. The minimization over
	$|\tr V_{m:1} (U_{m:1})^{-1}|$ then falls back to the even dimensional case, since the saturation case has a real trace.
\end{proof}

By using $\Phi(\mc V_{i}, \mc U_{i}) \approx \Phi(\mc A_{i}, \mc U_{i})/ {\Upsilon(\mc A_i)}$ we can formulate a quasi-saturated assessment about the average process fidelity of the circuit $\mc A_{m:1}$ given a partial information about its components $\mc A_i$ (in the equable scenario).
\begin{subequations}
	\begin{align}
	&{\text{For even dimensions:}}\notag \\
	&\cos^2\left(\sum\limits_{i=1}^m \arccos\left(\sqrt{\frac{\Phi(\mc A_{i}, \mc U_{i})}{{\Upsilon(\mc A_i)}}}\right)\right)  \prod\limits_{i=1}^m {\Upsilon(\mc A_i)}\lessapprox \Phi(\mc A_{m:1}, \mc U_{m:1}) \lessapprox \prod\limits_{i=1}^m {\Upsilon(\mc A_i)}~;\label{eq:phi_bounds_even}\\
	&{\text{for odd dimensions:}}\notag \\
	&\scalemath{0.8}{\left(\frac{(d-1)\cos\left(\sum_{i=1}^{m} \arccos\left(\frac{d \sqrt{\frac{\Phi(\mc A_{i}, \mc U_{i})}{{\Upsilon(\mc A_i)}}}-1}{d-1}\right)\right) +1}{d}\right)^2} \prod\limits_{i=1}^m {\Upsilon(\mc A_i)}\lessapprox \Phi(\mc A_{m:1}, \mc U_{m:1}) \lessapprox \prod\limits_{i=1}^m {\Upsilon(\mc A_i)}~.\label{eq:phi_bounds_odd}
	\end{align}
\end{subequations}
The terms in the cosine function are very close to what was defined as ``coherence angles'' in \cite{Dugas2016}.
Their sum can be interpreted as a coherent buildup. In some sense, the coherence angle is just another way to go about the notion of coherence level: it ties $r_{\rm coh}$ to an optimal rotation angle.

\subsection{Decoherence-limited operations}
When individual circuit elements $\mc A_i$ have purely decoherent 
equable errors, the bounds given by \cref{eq:phi_bounds_even,eq:phi_bounds_odd}
reduce to the approximate equality $\Phi(\mc A_{m:1}, \mc U_{m:1}) \approx \prod_i {\Upsilon(\mc A_i)}$. In fact, as long as the errors attached to the circuit elements $\mc A_i$ have a negligible
level of coherence, $\Phi(\mc A_{m:1}, \mc U_{m:1})$ is still expected to 
closely behave like 
a multiplicative decay. More rigorously, by looking more attentively 
at \cref{eq:phi_bounds_even,eq:phi_bounds_odd}, one should quickly realize that 
requiring 
\begin{align}\label{eq:decoh_limited}
	\Phi(\mc A_i, \mc U_i) = {\Upsilon(\mc A_i)}+O(r^2(\mc A_i, \mc U_i))
\end{align}
is sufficient to ensure
\begin{subequations}
	\begin{align}
	&\cos^2\left(\sum\limits_{i=1}^m \arccos\left(\sqrt{\frac{\Phi(\mc A_{i}, \mc U_{i})}{{\Upsilon(\mc A_i)}}}\right)\right) = 1+O(r^2(\mc A_{m:1}, \mc U_{m:1}))~, \\
	&{\text{and}}\notag \\
	&{\left(\frac{(d-1)\cos\left(\sum_{i=1}^{m} \arccos\left(\frac{d \sqrt{\frac{\Phi(\mc A_{i}, \mc U_{i})}{{\Upsilon(\mc A_i)}}}-1}{d-1}\right)\right) +1}{d}\right)^2} = 1+O(r^2(\mc A_{m:1}, \mc U_{m:1}))~.
	\end{align}
\end{subequations} 
A channel obeying the condition 
described by \cref{eq:decoh_limited} is said to be 
\emph{decoherence-limited}. The terminology is self-explanatory:
a channel is decoherence-limited if the infidelity to its target is
mostly limited by its decoherent infidelity $r_{\rm decoh}$, which cannot be
(substantially) reduced further through unitary corrections (see \cref{thm:max_cor}). Decoherence-limited
channels count decoherent channels, but also include channels for which
the infidelity of the coherent factor
plays a negligible role in the total infidelity, that is $r_{\rm coh}=O(r^2)$ or, equivalently, 
$r_{\rm coh}/r=O(r)$. 

Decoherent channels do not form a closed set under composition; the product 
of two positive semidefinite matrices is not necessarily positive semidefinite.
The geometric picture is that if two positive semidefinite contractions have different axes of contraction, they may induce (after composition) a small effective rotation.
However, the small rotation factor resulting from such composition is ensured to be 
very close to the identity, otherwise \cref{thm:fid_evo_2} wouldn't hold. 
More precisely, given two decoherent channels $\mc D_1$ and $\mc D_2$, 
the composite channel $\mc D_{2:1}= \mc V \circ \mc D'$ is such that 
$r(\mc V, \mc I)= O(r^2(\mc D',\mc I))$. In other words, 
the resulting channel is decoherence-limited. It is easy to see
from \cref{eq:phi_bounds_even,eq:phi_bounds_odd} that equable decoherence-limited
channels form a closed set under composition; if the coherence level of every
channel $\mc A_i$ in a circuit is of order $r(\mc A_i,\mc U_i)$, then
the coherence level of the total circuit is of order $r(\mc A_{m:1},\mc U_{m:1})$.

\subsection{Limitations}
In this section, we take a closer look at the bounds appearing
in \cref{thm:fid_evo,thm:uni_evo,thm:fid_evo_2,thm:uni_evo_2,thm:max_cor,thm:max_cor_multi,thm:fid_evo_3} and discuss their limitations.  
To parse through the expressions with more ease, consider $m$ channels $\mc A_i = \mc V_i \circ \mc D_i$ with identical
decoherent infidelity $r(\mc D_i, \mc I) = r_{\rm decoh}$. From this simplification, 
and by using $\Upsilon(\mc A_i) \approx \Upsilon(\mc A_i^\star) \approx \Phi(\mc D_i^\star,\mc I) \approx \Phi(\mc D_i, \mc I)$, which holds up to order $r_{\rm decoh}^2$, 
the margin of freedom in the bounds presented in this work reduces to the form\footnote{\Cref{thm:fid_evo_1} also contains a term of the form $r^2(\mc V_{m:1})$, but this term disappears in the equable regime.}
\begin{align}
	C_0~ m r^2_{\rm decoh} + C_1 m^2 r^2_{\rm decoh} +C_2~m r_{\rm decoh}~ r(\mc V_{m:1}, \mc I)  + \text{H.O.T.}~,\label{eq:leeway}
\end{align}
where $C_is$ are non-negative constants at most of order $1$ in the equable scenario. From \cref{thm:max_cor_multi}, the total infidelity scales at most as:
\begin{align}
	r(\mc A_{m:1}, \mc I) \lessapprox 1-\frac{d\left(1-\frac{d+1}{d}r_{\rm decoh}\right)^m+1}{d+1} = 
	m r_{\rm decoh} - \frac{d+1}{2d} m^2 r_{\rm decoh}^2~+ \text{H.O.T.}~,
\end{align} 
meaning that \cref{eq:leeway} is always at most of order $r^2(\mc A_{m:1}, \mc I)$. Hence, 
the bounds presented in this work apply very well in the high-fidelity regime. 

\begin{figure}[!b]
	\centering
	{\caption{
			Process fidelity of three different error channel compositions as a function of the circuit length. The hard lines correspond to the three process fidelities $\Phi(\mc A_{m:1}, \mc I)$, and the
			dotted lines correspond to the bounds given by \cref{thm:fid_evo_3}. The color map 
			illustrates the margin of freedom given by the RHS of \cref{eq:quasi_mult_U}, and the gray shaded area in the bottom plot corresponds to the limits of the top plot.
			In the top figure, shade variations indicate increments of $10^{-4}$, and in the bottom figure, increments of $10^{-2}$. The individual channels are of the form $\mc A_i = \mc V \circ \mc D$, where $\mc D$ is a dephasing channel with $\Phi(\mc D, \mc I) = 10^{-4}$, and $\mc V$ is a small unitary error. The dashed line is the decaying envelope $\Phi ^m(\mc D,\mc I)$. The three compositions differ by the level of coherence of their elements $\mc A_i$, which are $10 \%$, $1 \%$ and $0.01\%$. The lowest level of coherence corresponds to a decoherence-limited scenario, in which case the decoherent envelope stays within the bounds.		}\label{fig:saturation}}
	{\includegraphics[width=.9\textwidth]{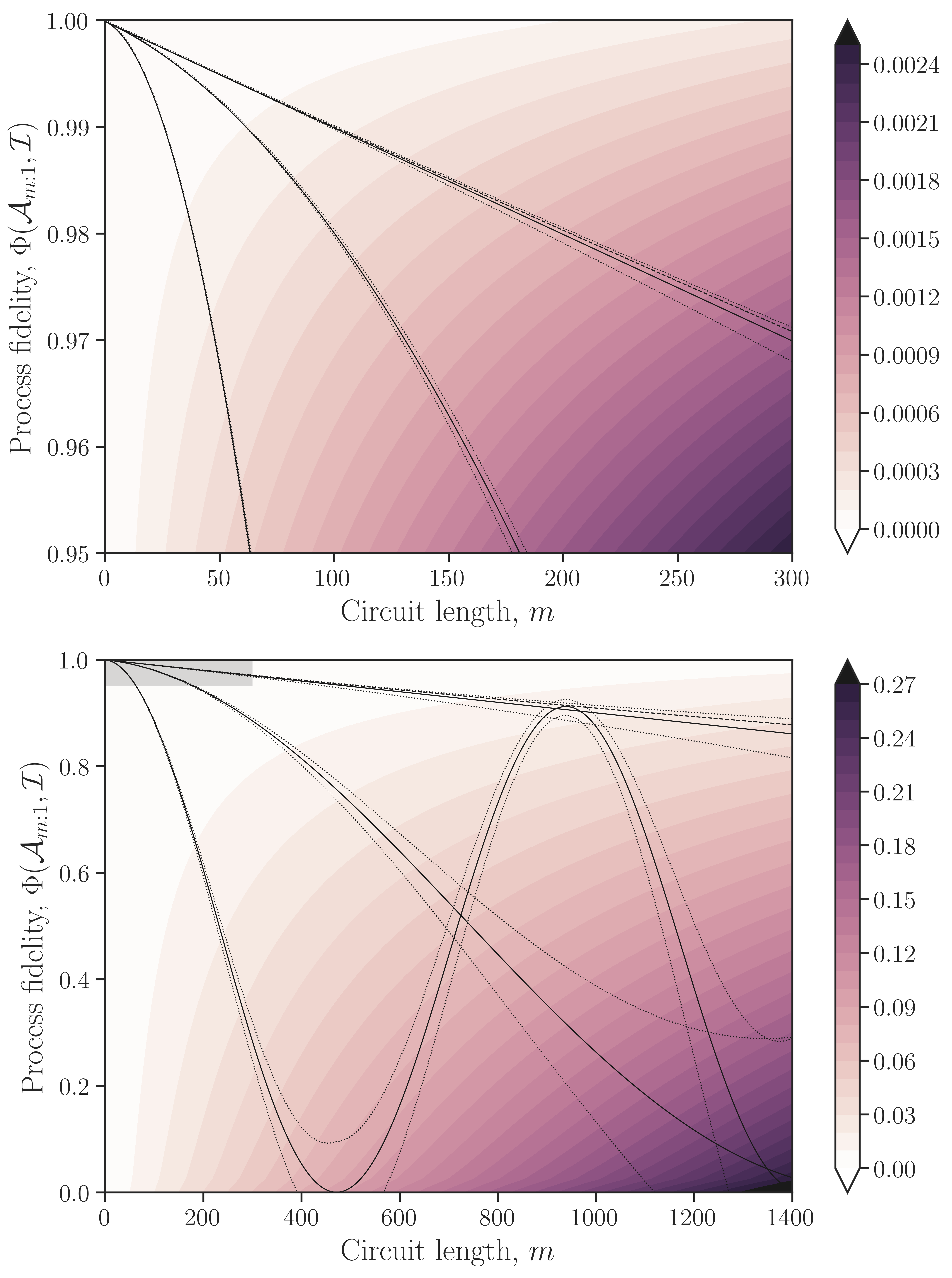}}
\end{figure}

As the fidelity decreases, the leeway portrayed by \cref{eq:leeway}
starts being noticeable. The appearance of quadratic terms of the form 
$m^2 r^2_{\rm decoh}$ is not surprising since most bounding techniques 
are based on the LK approximation, which ignores some $m^2 r^2_{\rm decoh}$ contributions.
To see this, consider $m$ identical channels $\mc A_i$ with canonical Kraus decomposition $\{ \sqrt{1-\delta} \mbb I, \sqrt{\delta} P \}$ where $\delta$ is small and $P$ is a unitary such that $P^2=\mbb I$ and $\tr P = 0$. Simple calculations yield
\begin{align}
	\Phi(\mc A_{m:1}^\star, \mc I) &= (1-\delta)^m = 1-m \delta + \frac{1}{2} m^2 \delta^2 + \text{H.O.T.} \\
	\Phi(\mc A_{m:1}, \mc I) &= \sum_{n=0}^{\lfloor m/2\rfloor} {m \choose {2n}} (1-\delta)^{m-2n} \delta^{2n} = 1-m \delta + \frac{1}{2} m^2 \delta^2+ {m \choose {2}}  \delta^{2} + \text{H.O.T.}
\end{align}
The term $\frac{1}{2} \left(\sum_i (1-\Upsilon(\mc A_i^\star))\right)^2$ featured in the 
bound of \cref{thm:fid_evo} is essentially achieved by the above example. However, 
not all the $m^2 r^2_{\rm decoh}$ terms appearing in the previous theorems are expected to 
be achieved by a composition of quantum channels. 

\Cref{fig:saturation} provides a good sense of the scaling of the bounds provided in \cref{thm:fid_evo_3}. In the figure, the decoherent infidelity of individual operations is of order $10^{-4}$. The top figure shows the bounds
for circuit lengths around $10^{2}$, in which case $m^2 r^2_{\rm decoh}$ is of order $10^{-4}$.
The bottom figure shows the bounds for circuit lengths around $10^{3}$, in which case $m^2 r^2_{\rm decoh}$ is of order $10^{-2}$. 
Once the circuit length $m$ is comparable to $r_{\rm decoh}^{-1}$ (in the example given by \cref{fig:saturation}, it would be as $m$ gets close to $10^{4}$), the fidelity is no longer ``small'', and $O(m^2 r_{\rm decoh}^2)$ becomes of order $1$, which renders the bounds trivial. In other words, to gain anything
valuable from the bounds in this work, the regime of consideration should be roughly 
$m^2 r^2_{\rm decoh} \lessapprox 10^{-1}$ and $m r_{\rm decoh} r(\mc V_{m:1}, \mc I) \lessapprox 10^{-1}$.
Notice that in such regime, as depicted by \cref{fig:saturation}, the only non-linear behavior in the composite fidelity must stem from unitary errors alone.




\section{Conclusion}
In this work, we investigated a
quasi-dynamical sub-parameterization of quantum channels that 
we referred to as the LK approximation. A remarkable realization is that 
this reduced picture still allows to
closely follow the evolution of two important figures of merit, namely the average process fidelity and the unitarity (see \cref{thm:uni_evo,thm:fid_evo}). 

Working with a simplified portrait sets aside superfluous subtleties 
and typically grants new mathematical properties to the object of consideration.
In our case, LK approximated mappings can be parameterized as contractions in $M_d(\mbb C)$; this set of matrices offers a much more intelligible categorization of error scenarios 
than the more abstruse full process matrix parameterization. 
Any matrix $A \in M_d(\mbb C)$ has a polar decomposition $V|A| $ where
$|A| \geq 0$ and $V$ is unitary. $V$ corresponds to a purely coherent 
physical operation $\mc V (\rho)= V \rho V^\dagger$, whereas the positive contraction $|A|$ 
is the LK operator belonging to what we classify as a decoherent channel (see \cref{defn:decoherent}).
In a nutshell, the polar decomposition in $M_d(\mbb C)$ translates into a 
coherent-decoherent factorization for quantum channels (see \cref{thm:decomp}).
We leveraged this dichotomy between types of noise
to derive fundamental principles of behavior concerning 
our two considered figures of merit. Among other properties, 
we demonstrated, up to high precision, the general monotonicity
of the unitarity as well as the monotonicity of the average process fidelity
of circuits with decoherent components (see \cref{thm:uni_evo_2,thm:fid_evo_1}). 

To pursue our analysis further, we introduced the wide-sense equable 
parameters $\gamma_{\rm decoh},\gamma_{\rm coh}$, which are defined through the LK parameterization (see \cref{defn:wse}). 
Equable error channels, for which $\gamma_{\rm decoh},\gamma_{\rm coh}$ are not too high, include all realistic noise models (and potentially more). 
Under the equability condition, we make multiple interesting 
connections between individual channels and compositions thereof:
\begin{enumerate}[i.]
		\item The infidelity of any channel can be 
		decomposed into a sum of two terms: a decoherent infidelity and a coherent one (respectively tied to the decoherent/coherent components of the channel). (See \cref{thm:fid_evo_3} and the discussion that immediately follows.)
		\item The unitarity, as well as the fidelity of circuits with decoherent elements, obey decay laws. Both these decays are closely dictated by the unitarity of individual components alone. (See \cref{thm:fid_evo_2,thm:max_cor,thm:uni_evo_2}.)
		\item The decoherent decay (that is, the decay prescribed by the decoherent factors of the circuit components) forms an upper bound to the total average process fidelity. Any substantial deviation from this upper bound is due to coherent effects alone (which gives us a lower bound). (See \cref{thm:max_cor_multi,thm:fid_evo_3}.)
\end{enumerate}

This work was primarily cast as a stepping-stone to formulate assessments about the performance of circuits based on partial knowledge of their constituents. While we do provide some assertion formulas, 
we want to emphasize that the more fundamental introduction of the LK 
approximation should also benefit the development of further characterization schemes. Indeed, the simple parameterization offered by the LK 
approximation facilitates the identification of specific noise signatures. 

\emph{Acknowledgments} --- The authors would like to thank Joel J. Wallman for his helpful discussions. This research was supported by the U.S. Army 
Research Office through grant W911NF-14-1-0103, TQT, CIFAR, the Government of 
Ontario, and the Government of Canada through CFREF, NSERC and Industry Canada.

\begin{table}[!h]
	\begin{minipage}{1.3\linewidth}
	\hspace{-0.3in}
	\fontten
	\begin{tabular}{|C{4cm}| C{4cm} |C{7cm}|}
		\hline
		Concept & Definition & Notes \\
		\hline 
		\hline 
		Non-catastrophic channel& $\Phi(\mc A, \mc U), \Upsilon^2(\mc A) > 1/2$ & 
		- Guarantees a unique LK operator. 
		\mbox{- Achieved given an acceptable level of control.}  \\
		\hline 
		LK operator, $A_1$ & Highest weight canonical Kraus operator, $A_1$  & 
		- Contains remarkable information about $\Phi$, $\Upsilon$. \\
		\hline
		LK approximation, $\mc A^\star$ & $\mc A^\star(\rho) = A_1 \rho A_1^\dagger$ & 
		- Replacing channels by their LK approximation in a circuit barely 
		affects its fidelity and unitarity. \\
		\hline  
		Decoherent channel & $A_1 \geq 0$& - Every non-catastrophic channel has a coherent-decoherent decomposition $\mc A = \mc U_A \circ \mc D_A = \mc D'_A \circ \mc U_A $.
		\mbox{- This definition of decoherence generalizes} the notion of decoherence in the Lindblad picture. \\
		\hline
		Extremal dephaser (channel) & $\exists~ \sigma_j \in \{\sigma_i(A_1)\}$ s.t. $1- \sigma_j \gg 1-\mbb E[\sigma_i]$ & - Strongly dephases a small set of states from the rest of the system. Since the set of states is small, extremal dephasers can still have high fidelity. \\
		\hline
		Extremal unitary (channel) & Let $\tr U \in \mbb R_+$.
		$\exists~ \lambda_j \in \{\lambda_i(U)\}$ s.t. $1- \text{Re}\{\lambda_j\} \gg 1-\mbb E[\text{Re}\{\lambda_i\}]$ & - Strongly dephases a small set of states from the rest of the system. Since the set of states is small, extremal dephasers can still have high fidelity. \\
		\hline
		WSE decoherence constant, $\gamma_{\rm decoh}$ & ${\rm SD}[\sigma_i] = \gamma_{\rm decoh} \mbb E[1-\sigma_i]$ & - For WSE channels,  $\gamma_{\rm decoh} \ll  1 / \sqrt{\mbb E[1-\sigma_i]}$. \\
		\hline
		WSE coherence constant $\gamma_{\rm coh}$ of unitary error $\mc U$ & Let $\tr U \in \mbb R_+$.
		$\text{SD}[\text{Re}\{\lambda_i\}] = \gamma_{\rm coh} \mbb E[1-\text{Re}\{\lambda_i\}]$ & -  For WSE channels, $\gamma_{\rm coh} \ll 1 / \sqrt{\mbb E[1- \text{Re}\{\lambda_i\}]}$.\\
		\hline  
		Equable channel & Non-catastrophic, no extremal errors (dephasers and unitaries). &
		\mbox{- Excludes pathological behaviors} \mbox{induced by extremal errors.}
		\mbox{- Should apply to all realistic scenarios.}
		\mbox{- Equable implies WSE.} \\
		\hline
		Wide-sense equable (WSE) channel & $\gamma_{\rm decoh} \ll  1 / \sqrt{\mbb E[1-\sigma_i]}~,$ 
		\mbox{$\gamma_{\rm coh} \ll 1 / \sqrt{\mbb E[1- \text{Re}\{\lambda_i\}]}$.} & \mbox{- Ensures the quasi-correspondence}: $\Phi(\mc D_A, \mc I) \approx {\Upsilon(\mc A)} \approx \max \limits_{W \in SU(d)} \Phi(\mc W \circ \mc A, \mc U) $
		\mbox{- Ensures the simple decay of the unitarity:} $ \Upsilon(\mc A_{m:1}) \approx \prod_i \Upsilon(\mc A_i) $
		\\
		\hline 
		Average gate fidelity, $F(\mc A, \mc U)$ & $\mbb E_{\rm Haar} f_{|\psi \rangle \langle \psi|}(\mc A, \mc U)$ & - Is the overlap between noisy and ideal outputs averaged over all physical inputs. \\
		\hline
		Unitarity, $u(\mc A)$ & $\mbb E_{\rm Haar} \frac{\|\mc A(|\psi \rangle \langle \psi|-\mbb I/d) \|_2^2}{\||\psi \rangle \langle \psi|-\mbb I/d \|_2^2}$& - Is the average contraction factor of the squared norm of the physical Bloch vectors. \\
		\hline
		$\Phi(\mc A, \mc U)$ &$\frac{(d+1)F(\mc A, \mc U) -1}{d}$ & - For non-catastrophic channels, $\Phi(\mc A_{m:1}, \mc U_{m:1}) \approx \Phi(\mc A^\star_{m:1}, \mc U_{m:1})$.
		\mbox{- For channels $\mc A_i =\mc V_i \circ \mc D_i  $ with WSE errors,} $\Phi(\mc A_{m:1}, \mc U_{m:1}) \approx \Phi(\mc V_{m:1}, \mc U_{m:1}) \prod_i \Phi(\mc D_{i}, \mc I)$~. \\
		\hline
		$\Upsilon^2(\mc A)$ & $\frac{(d^2-1)u(\mc A)+1}{d^2}$ &  - For non-catastrophic channels, $\Upsilon(\mc A_{m:1}) \approx \Upsilon(\mc A^\star_{m:1})$.
		\mbox{- In the WSE scenario,~~~~} $\Upsilon(\mc A_{m:1}) \approx \prod_i \Upsilon(\mc A_{i})$ ~. \\
		\hline
		Infidelity, $r(\mc A, \mc U)$ & $1-F(\mc A, \mc U)$ & \mbox{- For a channel $\mc A=\mc V \circ \mc D$,} (with WSE error)
		\mbox{~~~~~~~~~~~~~~~~~~~~$r = r_{\rm coh}+ r_{\rm decoh}+O(r^2)$~,~~~~~~~~~~~~~~~~~~~}
		~~where $r_{\rm coh}= r(\mc V, \mc U)$ and $r_{\rm decoh}= r(\mc D, \mc I)$.
		\\
		\hline
		Coherence level & $r_{\rm coh}/r$ & - Quantifies the proportion to which the error is coherent.\\
		\hline
		Decoherence-limited channel & $r_{\rm coh}/r = O(r)$ & - WSE decoherence-limited channels form a closed set under composition.\\
		\hline
	\end{tabular} 
	\caption{Summary of the main concepts addressed in this paper.}
\end{minipage}
\end{table}
\clearpage

\bibliography{../latex/library}{}
\clearpage
\appendix

\section{A noteworthy trace inequality}
This section is dedicated to demonstrating a useful trace inequality. 
\begin{boxlem}{Noteworthy trace inequality}{mat_ineq}
	Let $A, B \in M_d(\mbb C)$ be Hermitian matrices with eigenvalues of at most $\rho_A$, $\rho_B$ respectively.
	Then,
	\begin{align}
	\frac{\tr AB}{d} \geq \rho_B \frac{\tr A}{d} + \rho_A \frac{\tr B}{d} - \rho_A \rho_B~. \label{eq:matrix_ineq}
	\end{align}
\end{boxlem}
\begin{proof}
	We first show this inequality for positive semi-definite matrices
	with eigenvalues of at most $1$, under the condition that
	\begin{align}
	d < \lfloor \tr A \rfloor + \lfloor \tr B \rfloor +2~.
	\end{align}
	In such case, the inner product 
	is minimized by the sum of eigenvalues paired in opposite order \cite{WANG1993} (it's a matrix equivalent to the Hardy-Littlewood
	rearrangement inequality):
	\begin{align}
	\frac{\tr AB}{d} \geq \frac{1}{d} \sum_i \lambda_{i}^{\uparrow}(A) \lambda_{i}^\downarrow (B)~. \label{eq:rear_ineq}
	\end{align}
	This is in turn minimized when both $\{\lambda_i(A)\}$ and $\{\lambda_i(B)\}$ are maximized in terms of strong majorization. Since the eigenvalues are between zero and $1$, both majorizations have a simple form:
	\begin{align}
	\lambda_{i}(A)=&
	\begin{cases}
	1& \text{for}~ i \leq \lfloor \tr A \rfloor \\
	\tr A -\lfloor \tr A \rfloor & \text{for}~ i = \lfloor \tr A \rfloor+1\\
	0& \text{otherwise}~.
	\end{cases}  \\
	\lambda_{i}(B)= &
	\begin{cases}
	1& \text{for}~ i \leq \lfloor \tr B \rfloor \\
	\tr B -\lfloor \tr B \rfloor & \text{for}~ i = \lfloor \tr B \rfloor+1\\
	0& \text{otherwise}~.
	\end{cases} 
	\end{align}
	With such spectrum and the condition $ d < \lfloor \tr A \rfloor + \lfloor \tr B \rfloor +2$, we are ensured that
	\begin{align}
	\frac{1}{d} \sum_i \lambda_{i}^{\uparrow}(A) \lambda_{i}^\downarrow (B)= \frac{\tr A}{d} + \frac{\tr B}{d} -1~,
	\end{align}
	which, together with \cref{eq:rear_ineq}, yields \cref{eq:matrix_ineq} in this simpler case. 
	
	Now, consider the general case of Hermitian matrices $A,B$ with eigenvalues of at most $\rho_A, \rho_B$ respectively. 
	Let $A= (A+n_A\mbb I) - n_A \mbb I$, $B= (B+n_B\mbb I) - n_B \mbb I $, for $n_A, n_B \in \mbb R_+$, and consider the following expansion:
	\begin{align}
	\frac{\tr AB}{d} & = \frac{\tr (A+ n_A \mbb I)(B+ n_B \mbb I)}{d} - n_A\frac{\tr ( B+ n_B \mbb I)}{d} -n_B \frac{\tr( A+ n_A  \mbb I)}{d} +n_A n_B \notag \\
	& = (\rho_A+n_A)(\rho_B+n_B)\frac{\tr \left(\frac{A+ n_A \mbb I}{\rho_A+n_A}\right)\left(\frac{B+ n_B \mbb I}{\rho_B+n_B}\right)}{d} - n_A\frac{\tr B}{d} -n_B \frac{\tr  A }{d} - n_A n_B  \label{eq:expanded}
	\end{align}
	Now, let's pick $n_A, n_B$ large enough so that
	\begin{enumerate}[i.]
		\item $A+n_A \mbb I, B+n_B \mbb I \geq 0$~,
		\item $d < \lfloor \tr \left(\frac{A+ n_A \mbb I}{\rho_A+n_A}\right) \rfloor+\lfloor \tr \left(\frac{B+ n_B \mbb I}{\rho_B+n_B}\right) \rfloor +2$~.
	\end{enumerate}
	For $\rm i$ we can simply pick $n_A \geq \min \lambda(A)$, $n_B \geq \min \lambda(B)$. To see why $\rm ii$ is also possible, realize that
	\begin{align}
	\lim \limits_{n_A \rightarrow \infty} \left\lfloor \tr \left(\frac{A+ n_A \mbb I}{\rho_A+n_A}\right) \right\rfloor =d~,
	\end{align}
	meaning that there exists a finite $n_A$
	such that $\rm ii$ is fulfilled. Moreover, realize that the maximum eigenvalue
	of both $\frac{A+ n_A \mbb I}{\rho_A+n_A}$ and $\frac{B+ n_B \mbb I}{\rho_B+n_B}$ is upper-bounded by $1$ by construction. Combining all
	this, we get
	\begin{align}
	\frac{\tr \left(\frac{A+ n_A \mbb I}{\rho_A+n_A}\right)\left(\frac{B+ n_B \mbb I}{\rho_B+n_B}\right)}{d} \geq \frac{\tr \left(\frac{A+ n_A \mbb I}{\rho_A+n_A}\right)}{d}+\frac{\tr \left(\frac{B+ n_B \mbb I}{\rho_B+n_B}\right)}{d} -1~, \label{eq:simple_case}
	\end{align}
	since this corresponds to our initial simpler case. Substituting \cref{eq:simple_case} into \cref{eq:expanded} and simplifying, we get
	\cref{eq:matrix_ineq} which completes the proof.
\end{proof}
This inequality pairs well with the well-known Von-Neuman's trace inequality, as when $\tr AB \geq 0$, \cref{lem:mat_ineq} provides a much better lower bound. To see this, consider the following inequality which is trivially derived from Von's Neumann's trace inequality:
\begin{boxlem}{Flavored Von Neumann's trace inequality}{vn_mat_ineq}
	Let $A, B \in M_d(\mbb C)$ be matrices with spectral radius of at most $\rho_A$, $\rho_B$ respectively.
	Then,
	\begin{align}
	\left| \frac{\tr AB}{d} \right| \leq \min \left(\rho_B \frac{\tr |A|}{d},  \rho_A \frac{\tr |B|}{d}\right).
	\end{align}
	
\end{boxlem}

Recalling that $\|A\|_2^2 = \tr A^\dagger A$ and using those two last inequalities, we get the following norm inequality:
\begin{boxlem}{Norm inequality}{norm_ineq}
	Consider two matrices $A,B$ with spectral radius of at most $1$. Then,
	\begin{align}
	\frac{\|A\|^2_2}{d}+\frac{\|B\|^2_2}{d}-1 \leq \frac{\|AB\|^2_2}{d} \leq \min \left(\frac{\|A\|_2^2}{d},\frac{\|B\|_2^2}{d}\right)~.
	\end{align}    
\end{boxlem}

\section{Proofs of the main results}
\subsection{Notation and remarks}
Before we start proving \cref{thm:fid_evo,thm:uni_evo},
let's introduce some handy notation. The $i^{\rm th}$ canonical Kraus
operator of a channel $\mc A_j$ is denoted $A^{j}_{i}$.
Let $a\geq b$; we denote
\begin{align}
A^{a:b}_{\vec{i}} = A^{a}_{i_{a-b+1}} A^{a-1}_{i_{a-b}} \cdots A^{b+1}_{ i_2}A^{b}_{i_1}~,
\end{align}
where $\vec{i} \in \mbb N^{a-b+1}$
simply contains indices $i_k \in \{1, \cdots, d^2\}$. Finally we denote \mbox{$\vec{1} = (1,\cdots, 1)$} for which the dimension is left implicit.

Remark that the set $\{A^{m:1}_{\vec{i}}\}$
consist of a valid Kraus decomposition for the composite channel $\mc A_{m:1}$, and can be used to calculate $\Phi(\mc A_{m:1}, \mc U_{m:1})$ and $\Upsilon(\mc A_{m:1})$ through \cref{eq:chi00_expr,eq:calc_uni} respectively.
However, these Kraus operators are generally not orthogonal to one another (this is not the canonical decomposition), which prevents the same proof technique as in \cref{lem:uni,lem:fid}.

\subsection{Proof of the evolution \cref{thm:uni_evo}}
\begin{proof}
	Using H\"older's inequality, we get
	\begin{align}
	\Upsilon^2(\mc A_{m:1}) & = \sum \limits_{\vec{i}} \left(\frac{\|A^{m:1}_{\vec{i}}\|_2^2}{d}\right)^2 \label{eq:tot_uni
		_sum} \\
	& \leq \max \limits_{\vec{i}} \frac{\|A^{m:1}_{\vec{i}}\|_2^2}{d} \sum \limits_{\vec{j}} \frac{\|A^{m:1}_{\vec{j}}\|_2^2}{d} \tag{H\"older ineq.} \\
	& = \max \limits_{\vec{i}} \frac{\|A^{m:1}_{\vec{i}}\|_2^2}{d} \tag{TP condition}~.
	\end{align}
	One might have a (justified) hunch that 
	$ \underset{\vec{i}}{\rm{argmax}} \frac{\|A^{m:1}_{\vec{i}}\|_2^2}{d} 
	=\vec{1} $ in non-catastrophic noise scenarios. To show this, 
	consider $\vec{i}$ with $i_k \neq  1$ for some $k \in 
	\{1,\cdots ,m\}$. Using the properties of contractions, we have
	\begin{align}
	\frac{\|A^{m:1}_{\vec{i}}\|_2^2}{d} &\leq \frac{\|A^k_{i_k}\|_2^2}{d} \tag{Contractions} \\
	& \leq 1- \frac{\|A^k_{ 1}\|_2^2}{d} \tag{TP condition} \\
	&< 1/2 \tag{Non-catastrophic}~.
	\end{align}
	Hence, if we suppose $\underset{\vec{i}}{{\rm{argmax}}} \frac{\|A^{m:1}_{\vec{i}}\|_2^2}{d} \neq \vec{1}$, we have 
	\begin{align}
	\Upsilon^2(\mc A_{m:1}) < 1/2~,
	\end{align}
	which cannot be respected if the channel $\mc A_{m:1}$ is non-catastrophic. 
	Hence, by contradiction we have
	\begin{align}
	\Upsilon^2(\mc A_{m:1}) \leq \frac{\|A^{m:1}_{ 
			\vec{1}}\|_2^2}{d} = \Upsilon(\mc A_{m:1}^\star)~\label{eq:lower_bound_norm}.
	\end{align}
	From there we get
	\begin{align}
	\Upsilon^2(\mc A_{m:1}) & =\left(\frac{\|A^{m:1}_{ 
			\vec{1}}\|_2^2}{d}\right)^2 +  \sum \limits_{\vec{i}\neq \vec{1}} 
	\left(\frac{\|A^{m:1}_{ \vec{i}}\|_2^2}{d}\right)^2 \notag \\
	& \leq \Upsilon^2(\mc A^\star_{m:1}) + \left(\sum \limits_{\vec{i} \neq \vec{1}}
	\frac{\|A^{m:1}_{\vec{i}}\|_2^2}{d}\right)^2 \notag \\
	& = \Upsilon^2(\mc A^\star_{m:1}) + \left(1-
	\frac{\|A^{m:1}_{\vec{1}}\|_2^2}{d}\right)^2 \tag{TP condition} \\
	& = \Upsilon^2(\mc A^\star_{m:1}) + \left(1-\Upsilon(\mc A_{m:1}^\star)\right)^2\\
	& \leq \Upsilon^2(\mc A^\star_{m:1}) + \left(1-\Upsilon^2(\mc A_{m:1})\right)^2 ~\tag{\Cref{eq:lower_bound_norm}}~.
	\end{align}
\end{proof}
\subsection{Proof of the evolution \cref{thm:fid_evo}}
\begin{proof}
	We will show that the inequality \cref{eq:boundfidstar}
	holds for $m=2^n$, $\forall n \in \mbb N$. This suffices since if 
	$N < 2^n$, then we can append $\mc I_{2^n-N:1}$ to the composition $\mc A_{N:1}$
	so that $\mc A_{N:1} \circ \mc I_{2^n-N:1}$ is a composition of length $2^n$. Appending $\mc I_{2^n-N:1}$ has no effect on \cref{eq:boundfidstar}.
	
	From the definition of $\Phi$, we have that 
	$\Phi(\mc A_{m:1}, \mc U_{m:1})-\Phi(\mc A_{m:1}^\star, \mc U_{m:1}) \geq 0$, so it only remains to derive an upper bound on $\Phi(\mc A_{m:1}, \mc U_{m:1})-\Phi(\mc A_{m:1}^\star, \mc U_{m:1})$.  Our approach will be to split the sum as follows:
	\begin{align}
	\frac{1}{d^2}\sum \limits_{\vec{i} \neq \vec{1}} 
	\abs{\inner{A^{m:1}_{\vec{i}}}{U^{m:1}}}^2 = & 
	\frac{1}{d^2} \sum \limits_{\substack{ \vec{i}\neq \vec{1}}}
	\abs{\inner{A^{m:\frac{m}{2}+1}_{\vec{i}}A^{\frac{m}{2}:1}_{\vec{1}}}
		{U^{m:1}}}^2
	+\frac{1}{d^2}    \sum \limits_{\substack{ \vec{j}\neq \vec{1}}}
	\abs{\inner{A^{m:\frac{m}{2}+1}_{\vec{1}}A^{\frac{m}{2}:1}_{\vec{j}}}
		{U^{m:1}}}^2
	\notag     \\
	& + \frac{1}{d^2}    \sum \limits_{\substack{ \vec{i} \neq \vec{1} \\
			\vec{j}\neq  \vec{1}}}
	\abs{\inner{A^{m:\frac{m}{2}+1}_{\vec{i}}A^{\frac{m}{2}:1}_{\vec{j}}}
		{U^{m:1}}}^2~. \label{eq:split}
	\end{align}
	The double sum (last term) can be bounded via Cauchy-Schwarz inequality followed by the 
	usage of \cref{lem:norm_ineq}:
	\begin{align}
	\frac{1}{d^2}\sum \limits_{\substack{ \vec{i} \neq \vec{1} \\
			\vec{j}\neq  \vec{1}}}
	\abs{\inner{A^{m:\frac{m}{2}+1}_{\vec{i}}A^{\frac{m}{2}:1}_{\vec{j}}}
		{U^{m:1}}}^2~ &\leq  \sum \limits_{\substack{ \vec{i} \neq \vec{1} \\
			\vec{j}\neq  \vec{1}}} \frac{\|A^{m:\frac{m}{2}+1}_{ \vec{i}}\|_2^2}{d}\frac{\|A^{\frac{m}{2}:1}_{ \vec{j}}\|_2^2}{d} \tag{Cauchy-Schwarz ineq.} \\
	& \leq  \left(1-\frac{\|A^{m:\frac{m}{2}+1}_{ \vec{1}}\|_2^2}{d}\right)\left(1-\frac{\|A^{\frac{m}{2}:1}_{ \vec{1}}\|_2^2}{d}\right) \tag{TP condition} \\
	& \leq \left(\sum \limits_{i=\frac{m}{2}+1}^{m} \left(1-\frac{\|A^{i}_{ 1}\|_2^2}{d}\right)\right)\left(\sum \limits_{j=1}^{m/2} \left(1-\frac{\|A^{j}_{1}\|_2^2}{d}\right)\right) \tag{\Cref{lem:norm_ineq}}\\
	& \leq \left(\sum \limits_{i=\frac{m}{2}+1}^{m} (1-\Upsilon(\mc A_i^\star))\right)\left(\sum \limits_{j=1}^{m/2} (1-\Upsilon(\mc A_j^\star))\right) ~.
	\end{align}
	With regards to the first two terms on the RHS of \cref{eq:split}, let's 
	split them both into three terms once again:
	\begin{subequations}
		\begin{align}
		\frac{1}{d^2} \sum \limits_{\substack{ \vec{i}\neq \vec{1}}}
		\abs{\inner{A^{m:\frac{m}{2}+1}_{\vec{i}}A^{\frac{m}{2}:1}_{\vec{1}}}
			{U^{m:1}}}^2 =& 
		\frac{1}{d^2} \sum \limits_{\substack{ \vec{i}\neq \vec{1}}}
		\abs{\inner{A^{m:\frac{3m}{4}+1}_{\vec{i}}A^{\frac{3m}{4}:1}_{\vec{1}}}
			{U^{m:1}}}^2 \notag \\
		&+ \frac{1}{d^2} \sum \limits_{\substack{ \vec{i}\neq \vec{1}}}
		\abs{\inner{A^{m:\frac{3m}{4}+1}_{\vec{1}} A^{\frac{3m}{4}:\frac{m}{2}+1}_{\vec{j}} A^{\frac{m}{2}:1}_{\vec{1}}}
			{U^{m:1}}}^2 \notag     \\
		& + \frac{1}{d^2}    \sum \limits_{\substack{ \vec{i} \neq \vec{1} \\
				\vec{j}\neq  \vec{1}}}
		\abs{\inner{A^{m:\frac{3m}{4}+1}_{\vec{i}} A^{\frac{3m}{4}:\frac{m}{2}+1}_{\vec{j}} A^{\frac{m}{2}:1}_{\vec{1}}}
			{U^{m:1}}}^2~. \label{eq:splitagain1} \\
		\frac{1}{d^2} \sum \limits_{\substack{ \vec{i}\neq \vec{1}}}
		\abs{\inner{A^{m:\frac{m}{2}+1}_{\vec{1}}A^{\frac{m}{2}:1}_{\vec{i}}}
			{U^{m:1}}}^2 =  &
		\frac{1}{d^2} \sum \limits_{\substack{ \vec{i}\neq \vec{1}}}
		\abs{\inner{A^{m:\frac{m}{2}+1}_{ \vec{1}}A^{\frac{m}{2}:\frac{m}{4}+1}_{\vec{i}}A^{\frac{m}{4}:1}_{\vec{1}}}
			{U^{m:1}}}^2 
		\notag \\
		&+ \frac{1}{d^2} \sum \limits_{\substack{ \vec{j}\neq \vec{1}}}
		\abs{\inner{A^{m:\frac{m}{4}+1}_{\vec{1}}A^{\frac{m}{4}:1}_{\vec{j}}}
			{U^{m:1}}}^2 \notag     \\
		& + \frac{1}{d^2}    \sum \limits_{\substack{ \vec{i} \neq \vec{1} \\
				\vec{j}\neq  \vec{1}}}
		\abs{\inner{A^{m:\frac{m}{2}+1}_{\vec{1}}A^{\frac{m}{2}:\frac{m}{4}+1}_{\vec{i}}A^{\frac{m}{4}:1}_{\vec{j}}}
			{U^{m:1}}}^2~. \label{eq:splitagain2}
		\end{align}
	\end{subequations}
	The double sums on the RHS of \cref{eq:splitagain1,eq:splitagain2} can be upper bounded using the same technique
	as earlier, which yields
	\begin{align}
	&\frac{1}{d^2}    \sum \limits_{\substack{ \vec{i} \neq \vec{1} \\
			\vec{j}\neq  \vec{1}}}
	\abs{\inner{A^{m:\frac{3m}{4}+1}_{\vec{i}} A^{\frac{3m}{4}:\frac{m}{2}+1}_{\vec{j}} A^{\frac{m}{2}:1}_{\vec{1}}}
		{U^{m:1}}}^2+\frac{1}{d^2}    \sum \limits_{\substack{ \vec{i} \neq \vec{1} \\
			\vec{j}\neq  \vec{1}}}
	\abs{\inner{A^{m:\frac{m}{2}+1}_{\vec{1}}A^{\frac{m}{2}:\frac{m}{4}+1}_{\vec{i}}A^{\frac{m}{4}:1}_{\vec{j}}}
		{U^{m:1}}}^2 \notag \\
	&\leq \scalemath{0.95}{ \left(\sum \limits_{i=\frac{3m}{4}+1}^{m} (1-\Upsilon(\mc A_i^\star))\right)\left(\sum \limits_{j=\frac{m}{2}+1}^{3m/4} (1-\Upsilon(\mc A_j^\star))\right)+ \left(\sum \limits_{i=\frac{m}{4}+1}^{m/2} (1-\Upsilon(\mc A_i^\star))\right)\left(\sum \limits_{j=1}^{m/4} (1-\Upsilon(\mc A_j^\star))\right)}~.
	\end{align}
	By iterating the same subdivision technique, we end up with
	\begin{align}
	\frac{1}{d^2}\sum \limits_{\vec{i} \neq \vec{1}} 
	\abs{\inner{A^{m:1}_{ \vec{i}}}{U^{m:1}}}^2 &\leq \scalemath{0.95}{
	\sum \limits_{i=1}^{n} \sum\limits_{j=1}^{2^{i-1}} \left(\sum\limits_{k=\frac{2^n}{2^i}(2^i-2j+1)+1}^{\frac{2^n}{2^i}(2^i-2j+2)}(1-\Upsilon(\mc A_{k}^\star))\right)
	\left(\sum\limits_{k=\frac{2^n}{2^i}(2^i-2j)+1}^{\frac{2^n}{2^i}(2^i-2j+1)}(1-\Upsilon(\mc A_{k}^\star))\right)} \notag \\
	& +\frac{1}{d^2}\sum \limits_{\substack{j=1 \\ i \neq 1} }^{j=m} 
	\abs{\inner{A^{m:j+1}_{\vec{1}}A^{j}_{i}A^{j-1:1}_{\vec{1}}}{U^{m:1}}}^2 \label{eq:mostly_bounded}
	\end{align}
	Bounding the first term on the RHS can be done by alternating between the AM-GM inequality and
	square completions. First let's perform the AM-GM inequality on the terms of the summation restricted to $i=n$:
	\begin{align}\label{eq:using_AMGM}
	\sum\limits_{j=1}^{2^{n-1}} (1-\Upsilon(\mc A_{2^{n}-2j+2}^\star))
	(1-\Upsilon(\mc A_{2^{n}-2j+1}^\star)) \leq 
	\frac{1}{4}\sum\limits_{j=1}^{2^{n-1}} \left(\sum\limits_{k=2^{n}-2j+1}^{2^{n}-2j+2}(1-\Upsilon(\mc A_{k}^\star))\right)^2~. 
	\end{align}
	Then, let's add in the terms with index $i=n-1$ and complete the squares (taking $n=3$ as an example is recommended):
	\begin{align}
	&\sum \limits_{i=n-1}^{n} \sum\limits_{j=1}^{2^{i-1}} \left(\sum\limits_{k=\frac{2^n}{2^i}(2^i-2j+1)+1}^{\frac{2^n}{2^i}(2^i-2j+2)}(1-\Upsilon(\mc A_{k}^\star))\right)
	\left(\sum\limits_{k=\frac{2^n}{2^i}(2^i-2j)+1}^{\frac{2^n}{2^i}(2^i-2j+1)}(1-\Upsilon(\mc A_{k}^\star))\right)  \notag \\
	&\leq 
	\frac{1}{4}\sum\limits_{j=1}^{2^{n-1}} \left(\sum\limits_{k=2^{n}-2j+1}^{2^{n}-2j+2}(1-\Upsilon(\mc A_{k}^\star))\right)^2
	\notag \\
	~&+ 
	\sum\limits_{j=1}^{2^{n-2}} \left(\sum\limits_{k=2(2^{n-1}-2j+1)+1}^{2(2^{n-1}-2j+2)}(1-\Upsilon(\mc A_{k}^\star))\right)
	\left(\sum\limits_{k=2(2^{n-1}-2j)+1}^{2(2^{n-1}-2j+1)}(1-\Upsilon(\mc A_{k}^\star))\right)~, \tag{\cref{eq:using_AMGM}} \notag \\
	& \leq \frac{1}{4}\sum\limits_{j=1}^{2^{n-2}} \left(\sum\limits_{k=2(2^{n-1}-2j)+1}^{2(2^{n-1}-2j+2)}(1-\Upsilon(\mc A_{k}^\star))\right)^2 \notag \\
	&~+ 
	\frac{1}{2} \sum\limits_{j=1}^{2^{n-2}} \left(\sum\limits_{k=2(2^{n-1}-2j+1)+1}^{2(2^{n-1}-2j+2)}(1-\Upsilon(\mc A_{k}^\star))\right)
	\left(\sum\limits_{k=2(2^{n-1}-2j)+1}^{2(2^{n-1}-2j+1)}(1-\Upsilon(\mc A_{k}^\star))\right)\tag{Square completions} \\
	& \leq \frac{3}{8}\sum\limits_{j=1}^{2^{n-2}} \left(\sum\limits_{k=2(2^{n-1}-2j)+1}^{2(2^{n-1}-2j+2)}(1-\Upsilon(\mc A_{k}^\star))\right)^2~. \tag {AM-GM ineq.}
	\end{align}
	Similarly, we can then add in the terms with index $i=n-2$, complete the squares and
	use the AM-GM inequality on the leftover summation:
	\begin{align}
	&\sum \limits_{i=n-2}^{n} \sum\limits_{j=1}^{2^{i-1}} \left(\sum\limits_{k=\frac{2^n}{2^i}(2^i-2j+1)+1}^{\frac{2^n}{2^i}(2^i-2j+2)}(1-\Upsilon(\mc A_{k}^\star))\right)
	\left(\sum\limits_{k=\frac{2^n}{2^i}(2^i-2j)+1}^{\frac{2^n}{2^i}(2^i-2j+1)}(1-\Upsilon(\mc A_{k}^\star))\right) 
	\notag \\
	&\leq \frac{7}{16}\sum\limits_{j=1}^{2^{n-3}} \left(\sum\limits_{k=4(2^{n-2}-2j)+1}^{4(2^{n-2}-2j+2)}(1-\Upsilon(\mc A_{k}^\star))\right)^2~.
	\end{align}
	Repeating this procedure until $i=1$, we get
	\begin{align}
	&\sum \limits_{i=1}^{n} \sum\limits_{j=1}^{2^{i-1}} \left(\sum\limits_{k=\frac{2^n}{2^i}(2^i-2j+1)+1}^{\frac{2^n}{2^i}(2^i-2j+2)}(1-\Upsilon(\mc A_{k}^\star))\right)
	\left(\sum\limits_{k=\frac{2^n}{2^i}(2^i-2j)+1}^{\frac{2^n}{2^i}(2^i-2j+1)}(1-\Upsilon(\mc A_{k}^\star))\right) 
	\notag \\
	&~\leq \left(\frac{1}{2}- \frac{1}{2^{n+1}}\right) \left(\sum\limits_{i=1}^m (1-\Upsilon(\mc A_i^\star))\right)^2~. \label{eq:hardcore_ineq}
	\end{align}
	The last term on the RHS of \cref{eq:mostly_bounded} is upper-bounded using an alternate technique. First, we get
	\begin{align}
	\sum \limits_{\substack{j=1 \\ i \neq 1} }^{j=m} 
	\frac{\abs{\inner{A^{m:j+1}_{\vec{1}}A^{j}_{i}A^{j-1:1}_{\vec{1}}}{U^{m:1}}}^2}{d^2}&= 	\sum \limits_{\substack{j=1 \\ i \neq 1} }^{j=m} 
	\abs{\inner{A^{m:j+1}_{\vec{1}}\nlz{A}^{j}_{i}A^{j-1:1}_{\vec{1}}}{\nlz{U}^{m:1}}}^2 \cdot \frac{\|A^{j}_{i}\|_2^2}{d} \notag \\
	&\leq
	\max\limits_{\substack{j \\ i \neq 1} } \abs{\inner{A^{m:j+1}_{ \vec{1}}\nlz{A}^{j}_{i}A^{j-1:1}_{\vec{1}}}{\nlz{U}^{m:1}}}^2 
	\left(\sum \limits_{\substack{k=1 \\ \ell \neq 1} }^{k=m} \frac{\|A^{k}_{\ell}\|_2^2}{d}\right) ~
	\tag{H\"older's ineq.}    \\
	&=
	\max\limits_{\substack{j \\ i \neq 1} } \abs{\inner{\nlz{A}^{j}_{i}}{(A^{m:j+1}_{ \vec{1}})^\dagger\nlz{U}^{m:1}(A^{j-1:1}_{\vec{1}})^\dagger}}^2 ~
	\sum \limits_{\substack{k=1} }^{m}\left(1- \frac{\|A^{k}_{1}\|_2^2}{d}\right) 
	\tag{TP condition}\\
	& = \max\limits_{\substack{j \\ i \neq 1} } \abs{\inner{\nlz{A}^{j}_{i}}{(A^{m:j+1}_{ \vec{1}})^\dagger\nlz{U}^{m:1}(A^{j-1:1}_{\vec{1}})^\dagger}}^2
	~ \sum \limits_{\substack{k=1} }^{m}\left(1- \Upsilon(\mc A_k^\star)\right) \label{eq:some_ineq}~.
	\end{align}
	For fixed $j$, $\{\nlz{A}^{j}_{i}\}$ forms an orthonormal basis. Since 
	$\|(A^{m:j+1}_{\vec{1}})^\dagger\nlz{U}^{m:1}(A^{j-1:1}_{\vec{1}})^\dagger\|_2^2 \leq 1$ (contractions), we have that, for any $j$:
	\begin{align}
	\max\limits_{\substack{ i \neq 1} } \abs{\inner{\nlz{A}^{j}_{i}}{(A^{m:j+1}_{ \vec{1}})^\dagger\nlz{U}^{m:1}(A^{j-1:1}_{\vec{1}})^\dagger}}^2 
	&\leq 1- \abs{\inner{A^{m:j+1}_{ \vec{1}}\nlz{A}^{j}_{1}A^{j-1:1}_{\vec{1}}}{\nlz{U}^{m:1}}}^2 \notag \\
	& = 1- \frac{\abs{\inner{A^{m:j+1}_{ \vec{1}}A^{j}_{1}A^{j-1:1}_{\vec{1}}}{U^{m:1}}}^2}{d \|A^{j}_{1}\|_2^2} \notag  \\
	& = 1- \frac{\Phi(\mc A_{m:1}^\star, \mc U_{m:1})}{{\Upsilon(\mc A_j^\star)}}~\notag \\
	& \leq 1- \Phi(\mc A_{m:1}^\star, \mc U_{m:1})~.
	\label{eq:phi_ineq_max}
	\end{align}
	Combining \cref{eq:mostly_bounded,eq:hardcore_ineq,eq:some_ineq,eq:phi_ineq_max}
	yields
	\begin{align}
	\Phi(\mc A_{m:1}, \mc U_{m:1})-\Phi(\mc A_{m:1}^\star, \mc U_{m:1})&\leq \left(\frac{1}{2}- \frac{1}{2^{n+1}}\right) \left(\sum\limits_{i=1}^m (1-\Upsilon(\mc A_i^\star))\right)^2 \notag \\
	&~+(1- \Phi(\mc A_{m:1}^\star, \mc U_{m:1}))\sum \limits_{\substack{i=1} }^{m}\left(1- \Upsilon(\mc A_i^\star)\right)~.\label{eq:almost_there}
	\end{align}
	 To obtain obtain a lower bound that doesn't involve the LK approximation, we substitute $\Phi(\mc A^\star_{m:1}, \mc U_{m:1})$ by its
	lower bound, and $\Upsilon(\mc A_i^\star) \geq \Upsilon^2(\mc A_i)$:
	\begin{align}
	&{\Phi(\mc A_{m:1}, \mc U_{m:1})-\Phi(\mc A_{m:1}^\star, \mc U_{m:1})} \notag \\
	&\leq \frac{1}{2} \left(\sum\limits_{i=1}^m (1-\Upsilon^2(\mc A_i))\right)^2+(1- \Phi(\mc A_{m:1}, \mc U_{m:1}))\sum \limits_{\substack{i=1} }^{m}\left(1- \Upsilon^2(\mc A_i)\right) \notag \\
	& + \frac{1}{2} \left(\sum\limits_{i=1}^m (1-\Upsilon^2(\mc A_i))\right)^3+(1- \Phi(\mc A^\star_{m:1}, \mc U_{m:1}))\left(\sum \limits_{\substack{i=1} }^{m}\left(1- \Upsilon^2(\mc A_i)\right)\right)^2~. 
	\end{align}
\end{proof}    

\subsection{Proof of \cref{thm:fid_evo_2}}
The simplest route to prove theorems \ref{thm:uni_evo_2} to \ref{thm:fid_evo_3} is
probably to start with the demonstration of \cref{thm:fid_evo_2}.
\begin{proof}
	Given $m$ decoherent channels $\mc D_i$ with respective LK operators $D_1^i$, 
	we first want to bound the behavior of
	\begin{align}
	\sqrt{\Phi(\mc D_{m:1}^\star, \mc I)} = \frac{\tr \left(|D^m_{1}| \cdots |D^1_{1}|\right)}{d}~
	\end{align}
	as a function of the $\sqrt{\Phi(\mc D_{i}^\star, \mc I)}$s.
	Let's express the LK operators as \mbox{$|D^{i}_1|= \sqrt{\Phi(\mc D_{i}^\star, \mc I)}\mbb I_d + \Delta_i$}, and apply a telescopic expansion:
	\begin{align}
	\frac{\tr \left(|D^{m}_1| \cdots |D_{1}^1|\right)}{d} &= \prod_{i=1}^{m} \sqrt{\Phi(\mc D_{i}^\star, \mc I)} + \sum\limits_{j=1}^{m} \frac{\tr \left(|D_1^{m}| \cdots |D^{j+1}_1| \Delta_j\right)}{d}  \prod_{i=1}^{j-1} \sqrt{\Phi(\mc D_i^\star, \mc I)} \tag{Telescopic sum} \\
	& = \prod_{i=1}^{m} \sqrt{\Phi(\mc D_i^\star, \mc I)} + \left(\prod_{i=1}^{m} \sqrt{\Phi(\mc D_i^\star, \mc I)} \right) \sum\limits_{j=1}^{m} \frac{\tr \Delta_j}{d}/ \sqrt{\Phi(\mc D_j^\star, \mc I)} \notag \\
	&+\sum\limits_{j=1}^{m}\sum\limits_{k=j+1}^{m} \frac{\tr \left(|D^m_1| \cdots |D_1^{k+1}| \Delta_k \Delta_j \right)}{d} \prod_{\substack{i=1 \\ i \neq j}}^{k-1} \sqrt{\Phi(\mc D_i^\star, \mc I)}~. \tag{Telescopic sum, again} 
	\end{align}
	By construction, $\tr \Delta_i =0$, which leaves us with 
	\begin{align}
	\abs{\sqrt{\Phi(\mc D_{m:1}^\star, \mc I)}-\prod_{i=1}^{m} \sqrt{\Phi(\mc D_i^\star, \mc I)}} &\leq  \abs{\sum\limits_{j=1}^{m}\sum\limits_{k=j+1}^{m} \frac{\tr \left(|D^m_1| \cdots |D_1^{k+1}| \Delta_k \Delta_j \right)}{d} \prod_{\substack{i=1 \\ i \neq j}}^{k-1} \sqrt{\Phi(\mc D_i^\star, \mc I)}}~ \notag \\
	&\leq \sum\limits_{j=1}^{m}\sum\limits_{k=j+1}^{m} \abs{ \frac{\tr \left(|D^m_1| \cdots |D^{k+1}_1| \Delta_k \Delta_j \right) }{d}} \prod_{\substack{i=1 \\ i \neq j}}^{k-1} \sqrt{\Phi(\mc D_i^\star, \mc I)}~ \tag{Triangle ineq.} \\
	&\leq \sum\limits_{j=1}^{m}\sum\limits_{k=j+1}^{m} \abs{\frac{\tr \left(|D^m_1| \cdots |D_1^{k+1}| \Delta_k \Delta_j \right) }{d}} \tag{$\Phi(\mc D_i^\star, \mc I) \leq 1$} \\
	& \leq\sum\limits_{j=1}^{m}\sum\limits_{k=j+1}^{m}\frac{\| |D_1^m| \cdots |D_1^{k+1}| \Delta_k\|_2}{\sqrt{d}} \frac{\|\Delta_j \|_2 }{\sqrt{d}} \tag {Cauchy-Schwarz ineq.} \\
	& \leq\sum\limits_{j=1}^{m}\sum\limits_{k=j+1}^{m}\frac{\|\Delta_k\|_2}{\sqrt{d}} \frac{\|\Delta_j \|_2 }{\sqrt{d}} \tag {Contractions} 
	\end{align}
	This is where \cref{defn:wse} (equability) comes in handy, since it essentially states
	that $\frac{\|\Delta_i\|_2}{\sqrt{d}} = \gamma_{\rm decoh}(\mc D_i) \left(1-\sqrt{\Phi(\mc D_i^\star, \mc I)}\right)$. From there, we have
	\begin{align}
	\abs{\sqrt{\Phi(\mc D_{m:1}^\star, \mc I)}-\prod_{i=1}^{m} \sqrt{\Phi(\mc D_i^\star, \mc I)}}& \leq \gamma_{\rm decoh}^2 \sum\limits_{i=1}^{m}\sum\limits_{j=i+1}^{m}\left(1-\sqrt{\Phi(\mc D_j^\star, \mc I)}\right)\left(1-\sqrt{\Phi(\mc D_i^\star, \mc I)}\right) \label{eq:useful_for_thm_4} \\
	& \leq \gamma_{\rm decoh}^2 \sum\limits_{i=1}^{m}\sum\limits_{j=i+1}^{m}\left(1-\sqrt{\Phi(\mc D_j^\star, \mc I)}\right)\left(1-\sqrt{\Phi(\mc D_i^\star, \mc I)}\right) \notag \\
	& +\frac{\gamma_{\rm decoh}^2 }{2} \sum\limits_{i=1}^{m}\left(1-\sqrt{\Phi(\mc D_j^\star, \mc I)}\right)^2 \tag{Adding a positive term} \\
	& = \frac{\gamma_{\rm decoh}^2}{2}\left(\sum\limits_{i=1}^{m} \left(1-\sqrt{\Phi(\mc D_i^\star, \mc I)}\right) \right)^2 \label{eq:beaut} 
	\end{align}
	A few straightforward algebraic manipulations on \cref{eq:beaut} yield
	\begin{align}
	\abs{\Phi(\mc D_{m:1}^\star, \mc I) - \prod_{i=1}^{m}  \Phi(\mc D_i^\star,\mc I)} &\leq
	\gamma_{\rm decoh}^2 \prod_{i=1}^{m} \sqrt{\Phi(\mc D_i^\star,\mc I)} \left(\sum\limits_{i=1}^{m} \left(1-\sqrt{\Phi(\mc D_i^\star, \mc I)}\right) \right)^2
	\notag \\
	&+ ~ \frac{\gamma_{\rm decoh}^4}{4}\left(\sum\limits_{i=1}^{m} \left(1-\sqrt{\Phi(\mc D_i^\star, \mc I)}\right) \right)^4~. \label{eq:star_phi}
	\end{align}
	Using a simple telescopic expansion and \cref{lem:uni,thm:fid_evo}, we have
	\begin{align}
	\prod_{i=1}^{m}  \Phi(\mc D_i,\mc I)- \prod_{i=1}^{m}  \Phi(\mc D_i^\star,\mc I) &= \sum\limits_{j=1}^{m} \left(\prod_{i=j+1}^{m}  \Phi(\mc D_i,\mc I) \right) \left(\Phi(\mc D_j,\mc I)- \Phi(\mc D_j^\star,\mc I)\right) \left(\prod_{i=1}^{j-1}  \Phi(\mc D_i^\star,\mc I) \right) \notag \\
	&\leq \sum\limits_{i=1}^{m} \left(  1-\Upsilon(\mc D_i^\star) \right)\left(  1-\Phi(\mc D_i, \mc I) \right)~. \label{eq:star_mult}
	\end{align}
	From the triangle inequality we have
	\begin{align}
		\abs{\Phi(\mc D_{m:1}, \mc I) -\prod_{i=1}^{m}  \Phi(\mc D_i,\mc I)} &\leq \abs{\Phi(\mc D_{m:1}, \mc I) -\Phi(\mc D_{m:1}^\star, \mc I)}+\abs{\Phi(\mc D_{m:1}^\star, \mc I) - \prod_{i=1}^{m}  \Phi(\mc D_i^\star,\mc I)}\notag \\
		&~~+\abs{\prod_{i=1}^{m}  \Phi(\mc D_i^\star,\mc I)-\prod_{i=1}^{m} \Phi(\mc D_i,\mc I)}~.
	\end{align}
	Applying \cref{thm:fid_evo,eq:star_phi,eq:star_mult} on the RHS yields 
	\cref{eq:fid_evo_equ_dec}.
\end{proof}

\subsection{Proof of \cref{thm:uni_evo_2}}
\begin{proof}
	First, we derive an upper bound for $\Upsilon^2(\mc A_{m:1})$:
	\begin{align}
	\Upsilon^2(\mc A_{m:1}) &\leq 	\Upsilon^2(\mc A^\star_{m:1})+ (1-\Upsilon(\mc A^star_{m:1}))^2 \tag{\Cref{thm:uni_evo}}
	& = \left(\frac{\|A^{m:1}_{1}    \|_2^2}{d}\right)^2+ (1-\Upsilon(\mc A^\star_{m:1}))^2 \notag \\
	& \leq \min_i \left(\frac{\|A^i_{1}\|_2^2}{d}\right)^2+ (1-\Upsilon(\mc A^\star_{m:1}))^2 \tag{\Cref{lem:norm_ineq}} \\
	& \leq \min_i \Upsilon^2(\mc A_i)+ (1-\Upsilon(\mc A^\star_{m:1}))^2 \tag{\Cref{lem:uni}}~.
	\end{align}
	Before taking the square root on each side, notice that for any $\epsilon \leq 0$,
	the non-catastrophic condition enforces that $\sqrt{\Upsilon^2(\mc A_i)+ \epsilon} \leq \Upsilon(\mc A_i)+ \epsilon/ \sqrt{2}$. Indeed, since $\Upsilon(\mc A_i)> 1/\sqrt{2}$,
	\begin{align}
		& 1-\sqrt{2}\Upsilon(\mc A_i)<0 \notag \\
		\Rightarrow& \epsilon(1-\sqrt{2}\Upsilon(\mc A_i) - \epsilon/2)<0 \notag \\
		\Rightarrow& \Upsilon^2(\mc A_i)+\epsilon < \Upsilon^2(\mc A_i)+ \sqrt{2} \epsilon \Upsilon(\mc A_i) + \epsilon^2/2 \notag \\
		\Rightarrow& \sqrt{\Upsilon^2(\mc A_i)+\epsilon} < \Upsilon(\mc A_i)+ \epsilon/\sqrt{2}~.
	\end{align}
	Hence, 
	\begin{align}
		\Upsilon(\mc A_{m:1}) &\leq \min_i \Upsilon(\mc A_i)+ (1-\Upsilon(\mc A^\star_{m:1}))^2/\sqrt{2}~, \\
		&\min_i \Upsilon(\mc A_i)+ (1-\Upsilon^2(\mc A_{m:1}))^2/\sqrt{2}
	\end{align}	
	which corresponds to the quasi-monotonicity statement.
	We then derive a lower bound on $\Upsilon(\mc A_{m:1})$:
	\begin{align}
	1-\Upsilon(\mc A_{m:1}) &\leq 1-\Upsilon(\mc A^\star_{m:1}) \tag{\Cref{thm:uni_evo}} \\
	&= 1-\frac{\|A^{m:1}_{1}\|_2^2}{d} \notag \\
	& \leq \sum \limits_{i=1}^m \left(1-\frac{\|A^{i}_{1}\|_2^2}{d} \right)
	\tag{\Cref{lem:norm_ineq}} \\
	& \leq  \sum \limits_{i=1}^m \left(1-\sqrt{\Upsilon^2(\mc A_i)- (1-\Upsilon^2(\mc A_i))^2}\right) \tag{\Cref{lem:uni}}~.
	\end{align}
	Direct computation suffices to show that for $\Upsilon(\mc A_i) \in [2^{-1/2},1]$, 
	$\sqrt{\Upsilon^2(\mc A_i)- (1-\Upsilon^2(\mc A_i))^2}\leq \Upsilon(\mc A_i)-(1-\Upsilon^2(\mc A_i))^2$, hence
	\begin{align}
	1-\Upsilon(\mc A_{m:1})  \leq 
	\sum \limits_{i=1}^m \left(1-\Upsilon(\mc A_i)\right)+ (1-\Upsilon^2(\mc A_i))^2~, 
	\end{align}
	which corresponds to the quasi-subadditivity property.
	To derive the approximate multiplicativity statement, let's factor the decoherent channels 
	into their (left) polar decomposition $\mc A_i= \mc D_i \circ \mc V_i$. By relabeling
	$(\mc V_{i:1})^{-1}\circ \mc D_i \circ \mc V_{i:1} = \mc D'_i$, we have
	\begin{align}
	{\Upsilon(\mc A_{m:1}^\star)}
	&=\sqrt{\Phi(\mc {D'}_{1}^\star \cdots \mc {D'}_{m}^\star \mc {D'}_{m:1}^\star, \mc I)}~.
	\end{align}
	From \cref{defn:wse}, we have that ${\Upsilon(\mc {A}_i^\star)} - \Phi(\mc {D'}_i^\star,\mc I) \leq \gamma_{\rm decoh}^2 \left(1-\sqrt{\Phi(\mc {D'}_i^\star,\mc I)}\right)^2$. We can use a telescopic expansion to get
	\begin{align}
	\prod_i {\Upsilon(\mc {A}_i^\star)}- \prod_i \Phi(\mc {D'}_i^\star,\mc I)  
	&= \sum_{j=1}^{m} \left(\prod_{i=j+1}^{m} {\Upsilon(\mc {A}_i^\star)} \right)  
	({\Upsilon(\mc {A}_i^\star)} - \Phi(\mc {D'}_i^\star,\mc I)) \left(\prod_{i=1}^{j-1} \Phi(\mc {D'}_i^\star,\mc I)\right)
	\tag{Telescopic sum} \\
	&\leq 
	\gamma_{\rm decoh}^2 \sum_{j=1}^{m} \left(\prod_{i=j+1}^{m} {\Upsilon(\mc {A}_i^\star)} \right)  
	\left(1-\sqrt{\Phi(\mc {D'}_i^\star,\mc I)}\right)^2
	 \left(\prod_{i=1}^{j-1} \Phi(\mc {D'}_i^\star,\mc I)\right) ~
	\\
	&\leq
	\gamma_{\rm decoh}^2 \sum_{i=1}^{m} \left(1-\sqrt{\Phi(\mc {D'}_i^\star,\mc I)}\right)^2~ \notag \\
	&= 
	\gamma_{\rm decoh}^2 \sum_{i=1}^{m} \left(1-\sqrt{\Phi(\mc {D}_i^\star,\mc I)}\right)^2~
	.\label{eq:telescope} 
	\end{align}
	Using this, triangle inequality and \cref{eq:useful_for_thm_4}, 
	we get
	\begin{align}
	\abs{{\Upsilon(\mc {A}_{m:1}^\star)}-\prod_i {\Upsilon(\mc {A}_i^\star)}}
	&\leq \abs{\sqrt{\Phi(\mc {D'}_{1}^\star \cdots \mc {D'}_{m}^\star \mc {D'}_{m:1}^\star, \mc I)}-\prod_i \Phi(\mc {D'}_i^\star,\mc I)} \notag \\
	&~~+ \abs{\prod_i \Phi(\mc {D'}_i^\star,\mc I)- \prod_i {\Upsilon(\mc {A}_i^\star)}}
	\tag{Triangle ineq.}\\
	 &\leq 	4\gamma_{\rm decoh}^2 \sum\limits_{i=1}^{m}\sum\limits_{j=i+1}^{m}(1-\sqrt{\Phi(\mc {D'}_j^\star, \mc I)})(1-\sqrt{\Phi(\mc {D'}_i^\star, \mc I)}) 
	\notag  \\&~~+  	
	2\gamma_{\rm decoh}^2 \sum_i \left(1-\sqrt{\Phi(\mc {D'}_i^\star,\mc I)}\right)^2
	\tag{\Cref{eq:useful_for_thm_4,eq:telescope}} \\
	&= 2 \gamma_{\rm decoh}^2 \left(\sum_{i=1}^m \left(1-\sqrt{\Phi(\mc {D'}_i^\star,\mc I)}\right)\right)^2~ \tag{Complete the square} \\
	&= 2 \gamma_{\rm decoh}^2 \left(\sum_{i=1}^m \left(1-\sqrt{\Phi(\mc {D}_i^\star,\mc I)}\right)\right)^2~ .
	\label{eq:uni_before_sq}
	\end{align}
	Notice that a usage of \cref{thm:uni_evo} allows to translate $\Upsilon(\mc A_{m:1}^\star)$
	into $\Upsilon(\mc A_{m:1})$:
	\begin{align}
	(1-\Upsilon(\mc A^\star_{m:1}))^2 &\geq \Upsilon^2(\mc A_{m:1})- \Upsilon^2(\mc A_{m:1}^\star) \tag{\Cref{thm:uni_evo}} \\
	& = \left( \Upsilon(\mc A_{m:1})- \Upsilon(\mc A_{m:1}^\star)\right)\left( \Upsilon(\mc A_{m:1})+ \Upsilon(\mc A_{m:1}^\star)\right) \notag \\
	& >  \Upsilon(\mc A_{m:1})- \Upsilon(\mc A_{m:1}^\star) \tag{Non-catastrophic condition}~.
	\end{align}
	To remove the LK approximations from $\prod_i \Upsilon(\mc {A}_i^\star)$, we use
	\begin{align}
	\prod_i \Upsilon(\mc {A}_i)- \prod_i \Upsilon(\mc {A}_i^\star) &= \sum_{j=1}^m \prod_{i=j+1}^{m}\Upsilon(\mc {A}_i) (\Upsilon(\mc {A}_j)- \Upsilon(\mc {A}_j^\star))\prod_{i=1}^{j-1}\Upsilon(\mc {A}_i^\star)  \\
	&\leq \sum_{j=1}^m (\Upsilon(\mc {A}_j)- \Upsilon(\mc {A}_j^\star)) \\
	&\leq \sum_{j=1}^m  (1- \Upsilon(\mc {A}^\star_j))^2~. \label{eq:bound_products}
	\end{align}
	Using the triangle inequality and \cref{eq:telescope,eq:uni_before_sq,eq:bound_products,thm:uni_evo} yields
	\begin{align}
	\abs{\Upsilon(\mc A_{m:1})- \prod_i \Upsilon(\mc {A}_i)} &\leq 
	(1- \Upsilon(\mc A_{m:1}^\star))^2 +  \sum_{j=1}^m  (1- \Upsilon(\mc {A}^\star_j))^2 \notag \\
	&+\gamma_{\rm decoh}^2 \sum_{i=1}^{m} \left(1-\sqrt{\Phi(\mc {D}_i^\star,\mc I)}\right)^2 \notag \\
	&+2 \gamma_{\rm decoh}^2 \left(\sum_{i=1}^m \left(1-\sqrt{\Phi(\mc {D}_i^\star,\mc I)}\right)\right)^2
	\end{align}
	Invoking \cref{thm:max_cor} allows to naturally translates between 
	$\Phi(\mc D_i^\star,\mc I)$ and $\Upsilon(\mc A_i)$, which completes the proof.
	
\end{proof}

\subsection{Proof of \cref{thm:max_cor}}

\begin{proof}
	Let $\sigma_i$ be the singular values of the LK operator of $\mc A$.
	The first part of the proof revolves around
	\begin{align}
	\mbb E (\sigma_i^2) \leq \mbb E (\sigma_i ) \leq \sqrt{\mbb E (\sigma_i^2)}~,\label{eq:sigm_id} 
	\end{align}
	which implies that 
	\begin{align}
	\Upsilon^2(\mc A^\star) \leq \left(\mbb E (\sigma_i)\right)^2 \leq {\Upsilon(\mc A^\star)}~\label{eq:uni_fid_id}~. 
	\end{align}
	First, let's demonstrate the lower bound \cref{eq:lower_phi_corr}:
	\begin{align}
	\max\limits_{V \in SU(d)} \Phi(\mc V \circ \mc A, \mc U) &\geq \max\limits_{V \in SU(d)} \Phi(\mc V \circ \mc A^\star, \mc U) \tag{\Cref{thm:fid_evo}} \\
	& = \left(\mbb E (\sigma_i)\right)^2 \label{eq:some_line} \\
	& \geq \Upsilon^2(\mc A^\star) \tag{\Cref{eq:uni_fid_id}} \\
	& \geq \Upsilon^2(\mc A)- (1-\Upsilon^2(\mc A))^2 \tag{\Cref{lem:uni}}~.
	\end{align}
	Demonstrating the upper bound \cref{eq:upper_phi_corr} follows the same reasoning:
	\begin{align}
	\max\limits_{V \in SU(d)} \Phi(\mc V \circ \mc A, \mc U) &\leq \Upsilon(\mc A^\star)\max\limits_{V \in SU(d)} \Phi(\mc V \circ \mc A^\star, \mc U)+(1-\Upsilon(\mc A^\star))+ \frac{1}{2}(1-\Upsilon(\mc A^\star))^2
	 \tag{\Cref{eq:almost_there}} \\
	& = \Upsilon(\mc A^\star)\left(\mbb E (\sigma_i)\right)^2 + (1-\Upsilon(\mc A^\star))+ \frac{1}{2}(1-\Upsilon(\mc A^\star))^2 \notag \\
	& \leq  {\Upsilon^2(\mc A^\star)} + (1-\Upsilon(\mc A^\star))+ \frac{1}{2}(1-\Upsilon(\mc A^\star))^2~\tag{\Cref{eq:uni_fid_id}} 
	\\
	& =  {\Upsilon(\mc A^\star)} + \frac{3}{2}(1-\Upsilon(\mc A^\star))^2 \notag \\
	& \leq  {\Upsilon(\mc A)} + \frac{3}{2}(1-\Upsilon^2(\mc A))^2  \tag{\Cref{lem:uni,eq:lower_bound_norm} }~.
	\end{align}
	To tighten the lower bound at line \ref{eq:some_line}, we may use the WSE decoherence constant:
	\begin{align}
	\left(\mbb E (\sigma_i)\right)^2 &= {\Upsilon(\mc A^\star)}-\gamma_{\rm decoh}^2 (1-\mbb E(\sigma_i))^2 \tag{Equability}
	\\
	&\geq {\Upsilon(\mc A^\star)} - \gamma_{\rm decoh}^2 (1-{\Upsilon(\mc A^\star)})^2\tag{\Cref{eq:uni_fid_id}} \\
	& \geq {\Upsilon(\mc A)-(1-\Upsilon(\mc A^\star))^2} - \gamma_{\rm decoh}^2 (1-{\Upsilon(\mc A^\star)})^2 \tag{\Cref{lem:uni}} \\
	& \geq {\Upsilon(\mc A)}-(1+\gamma_{\rm decoh}^2)(1-\Upsilon^2(\mc A))^2~, \tag{($\Upsilon^2(\mc A) \leq \Upsilon(\mc A^\star)$)} 
	\end{align}
	which completes the proof.
\end{proof}

\subsection{Proof of \cref{thm:fid_evo_1}}    
\begin{proof}
	First, we derive an upper bound for $\Phi(\mc V \circ \mc D_{m:1}, \mc I)$:
	\begin{align}
	\Phi(\mc V \circ \mc D_{m:1}, \mc I) &\leq \Phi(\mc V \circ \mc D_{m:1}^\star, \mc I) \notag \\
	&~~+  \frac{1}{2} \left(\sum\limits_{i=1}^m (1-\Upsilon(\mc D_i^\star))\right)^2+
	(1- \Phi(\mc V\circ \mc D_{m:1}^\star, \mc I))\sum \limits_{\substack{i=1} }^{m}\left(1- \Upsilon(\mc D_i^\star)\right)~. \tag{\Cref{thm:fid_evo}}
	\end{align}
	Using \cref{lem:vn_mat_ineq,lem:fid}, we get (let $D_i^j$ be the $i^{\rm th}$ canonical Kraus operator of $\mc D_j$)
	\begin{align}
	\Phi(\mc V \circ \mc D_{m:1}^\star, \mc I) & \leq \abs{\frac{\tr \left( V |D_1^m| \cdots |D_1^1| \right) }{d}}^2\notag \\
	& \leq \min_i \abs{\frac{\tr |D^i_1|}{d}}^2 \tag{\Cref{lem:vn_mat_ineq}} \\
	& \leq \min_i \Phi(\mc D_i,\mc I) \tag{\Cref{lem:fid}} ~,
	\end{align}
	which yields the quasi-monotonicity statement. Now, we derive a lower bound for \mbox{$\Phi(\mc V \circ \mc D_{m:1}, \mc I)$}:
	\begin{align}
	1-\sqrt{\Phi(\mc V \circ  \mc D_{m:1}, \mc I)} &\leq 1-\sqrt{\Phi(\mc V \circ\mc D_{m:1}^\star, \mc I)} \tag{\Cref{thm:fid_evo}} \\
	& = 1- \abs{\frac{\tr \left( V |D_1^m| \cdots |D_1^1|\right)}{d}} 
	\end{align}
	At this point, it seems tempting to use \cref{lem:mat_ineq}, but recall that $V$
	is generally not Hermitian. However, we can get by as follows
		\begin{align}
	1-\sqrt{\Phi(\mc V \circ  \mc D_{m:1}, \mc I)} &\leq
	 1- \abs{{\text{Re}} \left\{\frac{\tr \left(V |D_1^m| \cdots |D_1^1|\right)}{d}\right\}} \notag \\
	& = 1- \abs{\frac{\tr \left({\text{Re}} (V) |D_1^m| \cdots |D_1^1|\right)}{d}}~,
	\end{align}
	where ${\text{Re}}(V):=(V+V^\dagger)/2$ is Hermitian, which allows us to use
	\cref{lem:mat_ineq}:
	\begin{align}
    	1-\sqrt{\Phi(\mc V \circ  \mc D_{m:1}, \mc I)}	& \leq \left(1- \abs{\frac{\tr {\text{Re}}(V)}{d}}\right) +\sum\limits_{i=1}^m \left(1- \abs{\frac{\tr |D_1^i|}{d}}\right) \tag{\Cref{lem:mat_ineq}} 
	\end{align}
	WOLOG, we pick the global phase of $V$
	such that $\sqrt{\Phi(\mc V, \mc I)}= \tr V /d \in \mbb R_+$. From
	there we get
	\begin{align}
	1-\sqrt{\Phi(\mc V \circ  \mc D_{m:1}, \mc I)}	& \leq \left(1- \sqrt{\Phi(\mc V, \mc I)}\right) + \sum\limits_{i=1}^m \left(1- \sqrt{\Phi(\mc D_i^\star, \mc I)}\right)~.
	\end{align}
	 To remove the square roots and the star, let's use $ 1-x/2- x^2/ 2\leq \sqrt{1-x} \leq 1-x/2$
	for $x \in [0,1]$ and \cref{lem:fid}:
	\begin{align}
	1-\Phi(\mc V \circ \mc D_{m:1}, \mc I)&\leq \left(1- {\Phi(\mc V, \mc I)}\right)+ \sum\limits_{i=1}^m (1-\Phi(\mc D_i, \mc I)) \notag \\
	&+ (1- \Phi(\mc V, \mc I))^2+ \sum\limits_{i=1}^m (1- \Phi(\mc D_i^\star, \mc I))^2+\sum\limits_{i=1}^m (1- \Phi(\mc D_i, \mc I))(1-\Upsilon^2(\mc D_i)) 
	\end{align}
	which corresponds to the quasi-subadditivity property.
\end{proof}    

\subsection{Proof of \cref{thm:fid_evo_3}}
\begin{proof}
Our goal is to bound 
\begin{align}
	\abs{\frac{\tr \left( V |D^m_1|\cdots|D_1^1| \right) }{d}}^2 = \abs{\text{Re}\left\{\frac{\tr \left(V |D^m_1|\cdots|D_1^1|\right) }{d}\right\}}^2 +\abs{\text{Im}\left\{\frac{\tr \left(V |D^m_1|\cdots|D_1^1| \right)}{d}\right\}}^2~.
\end{align}
Proving \cref{thm:fid_evo_3} is very similar to proving
\cref{thm:fid_evo_2}, but the appended unitary $V$ requires some extra care.
Let's first bound the amplitude of the imaginary term. 
WOLOG, we pick the global phase of $V$
such that $\sqrt{\Phi(\mc V, \mc I)}= \tr V /d \in \mbb R_+$.
\begin{align}
	\abs{\text{Im}\left\{\frac{\tr \left(V |D^m_1|\cdots|D_1^1| \right)}{d}\right\}}  
	& = \abs{\text{Im}\left\{\tr \left[ \frac{(V-\tr (V)/d\mbb ~\mbb I) (|D^m_1|\cdots|D_1^1|-\tr (|D^m_1|\cdots|D_1^1|) /d ~\mbb I )}{d}\right]\right\}} \tag{Adding real terms.} \\
	& \leq  \abs{\tr \left[\frac{(V-\tr (V)/d\mbb ~\mbb I) (|D^m_1|\cdots|D_1^1|-\tr (|D^m_1|\cdots|D_1^1|) /d ~\mbb I )}{d} \right]} \notag \\
	&\leq  \frac{\|V-\tr (V)/d\mbb I\|_2}{\sqrt d} \frac{\| |D^m_1|\cdots|D_1^1|-\tr (|D^m_1|\cdots|D_1^1|)/d \mbb I\|_2}{\sqrt d}  \tag{Cauchy-Schwarz ineq.} \\
	&=  \sqrt{1-\Phi(\mc{V}, \mc{I})}\sqrt{{\Upsilon(\mc D_{m:1}^\star)}- \Phi(\mc D_{m:1}^\star, \mc{I})} ~.
\end{align}
We know from \cref{thm:uni_evo_2} that ${\Upsilon(\mc D_{m:1}^\star)} \approx \prod_i {\Upsilon(\mc D_{i}^\star)}$. We also know from \cref{thm:max_cor} that 
$ {\Upsilon(\mc D_{i}^\star)} \approx \Phi(\mc D_i^\star,\mc I)$. From 
\cref{thm:fid_evo_2} we know that $\prod_i \Phi(\mc D_i^\star,\mc I) \approx \Phi(\mc 
D_{m:1}^\star,\mc I)$. By combining this information, we have that ${\Upsilon(\mc 
D_{m:1}^\star)} \approx \Phi(\mc D_{m:1}^\star, \mc{I})$. More precisely, by using
 \cref{eq:star_phi,eq:telescope,eq:uni_before_sq}, we get
\begin{align}
	&\abs{\text{Im}\left\{\frac{\tr \left(V |D^m_1|\cdots|D_1^1| \right)}{d}\right\}}^2 \leq(1-\Phi(\mc{V}, \mc{I})) \Bigg[\gamma_{\rm decoh}^2 \prod_{i=1}^{m} \sqrt{\Phi(\mc D_i^\star,\mc I)} \left(\sum\limits_{i=1}^{m} \left(1-\sqrt{\Phi(\mc D_i^\star, \mc I)}\right) \right)^2
	\notag \\
	&~~~~~~~+  \frac{\gamma_{\rm decoh}^4}{4}\left(\sum\limits_{i=1}^{m} \left(1-\sqrt{\Phi(\mc D_i^\star, \mc I)}\right) \right)^4
	+\gamma_{\rm decoh}^2 \sum_{i=1}^{m} \left(1-\sqrt{\Phi(\mc {D}_i^\star,\mc I)}\right)^2 \notag \\ &
	~~~~~~~+ 2 \gamma_{\rm decoh}^2 \left(\sum_{i=1}^m \left(1-\sqrt{\Phi(\mc {D}_i^\star,\mc I)}\right)\right)^2 \Bigg] ~,
\end{align}
meaning that the imaginary term is absolutely insignificant. To bound the real part of the trace, we
mimic most of the proof technique used to prove \cref{thm:fid_evo_2}. Let's express the LK operators as $|D^{i}_1|= \sqrt{\Phi(\mc D_{i}^\star, \mc I)}\mbb I_d + \Delta_i$ and $V=\sqrt{\Phi(\mc V, \mc I)}\mbb I_d + \Delta_{m+1}$, and apply a first telescopic expansion:
\begin{align}
	\text{Re} \left\{\tr \left[\frac{ V|D^{m}_1| \cdots |D_{1}^1|}{d} \right]\right\}&=\sqrt{\Phi(\mc V, \mc I)} \prod_{i=1}^{m} \sqrt{\Phi(\mc D_{i}^\star, \mc I)} \notag \\
	&~~+ \sum\limits_{j=1}^{m}\text{Re} \left\{ \tr \left[ \frac{ V|D_1^{m}| \cdots |D^{j+1}_1| \Delta_j}{d} \right] \right\}\prod_{i=1}^{j-1} \sqrt{\Phi(\mc D_i^\star, \mc I)} \tag{Telescopic sum} 
	\end{align}
	By applying the expansion again, and use $\tr \Delta_i =0$, we get:
	\begin{align}
	\text{Re} \left\{\tr \left[ \frac{ V|D^{m}_1| \cdots |D_{1}^1|}{d} \right]\right\}& = \sqrt{\Phi(\mc V, \mc I)} \prod_{i=1}^{m} \sqrt{\Phi(\mc D_{i}^\star, \mc I)}  \notag \\
	&~~+\sum\limits_{j=1}^{m}\sum\limits_{k=j+1}^{m} \text{Re} \left\{ \frac{\tr \left( V|D^m_1| \cdots |D_1^{k+1}| \Delta_k \Delta_j \right)}{d} \right\} \prod_{\substack{i=1 \\ i \neq j}}^{k-1} \sqrt{\Phi(\mc D_i^\star, \mc I)} \notag \\
	&~~+ 
	\sum\limits_{j=1}^{m} \text{Re} \left\{ \frac{\tr \left(\Delta_{m+1} \Delta_j \right)}{d} \right\} \prod_{\substack{i=1 \\ i \neq j}}^{m} \sqrt{\Phi(\mc D_i^\star, \mc I)}
	~. \tag{Telescopic sum, again} 
\end{align}
After a simple application of the triangle inequality, we get
\begin{align}
	&\abs{\text{Re} \left\{\tr \left[\frac{ V|D^{m}_1| \cdots |D_{1}^1|}{d} \right]\right\}-\sqrt{\Phi(\mc V, \mc I)} \prod_{i=1}^{m} \sqrt{\Phi(\mc D_{i}^\star, \mc I)} } \leq  \abs{\sum\limits_{j=1}^{m} \text{Re} \left\{ \frac{\tr \left(\Delta_{m+1} \Delta_j \right) }{d} \right\} \prod_{\substack{i=1 \\ i \neq j}}^{m} \sqrt{\Phi(\mc D_i^\star, \mc I)}} \notag \\
	& ~~~~~~~~~~~~~~~~~~~~~~~~~~~~~~~~~~~~~~~~
	+  \abs{\sum\limits_{j=1}^{m}\sum\limits_{k=j+1}^{m} \text{Re} \left\{ \frac{\tr \left(V|D^m_1| \cdots |D_1^{k+1}| \Delta_k \Delta_j \right)}{d} \right\} \prod_{\substack{i=1 \\ i \neq j}}^{k-1} \sqrt{\Phi(\mc D_i^\star, \mc I)}} \label{eq:real_stuff}
\end{align}
The second term on the RHS is upper-bounded by the exact same technique as in \cref{thm:fid_evo_2} (see the derivation of \cref{eq:beaut}):
\begin{align}
	\scalemath{0.9}{\abs{\sum\limits_{j=1}^{m}\sum\limits_{k=j+1}^{m} \text{Re} \left\{ \frac{\tr \left(V|D^m_1| \cdots |D_1^{k+1}| \Delta_k \Delta_j\right) }{d} \right\} \prod_{\substack{i=1 \\ i \neq j}}^{k-1} \sqrt{\Phi(\mc D_i^\star, \mc I)}} \leq \frac{\gamma_{\rm decoh}^2}{2}\left(\sum\limits_{i=1}^{m} \left(1-\sqrt{\Phi(\mc D_i^\star, \mc I)}\right) \right)^2}~.
\end{align}
The first term on the RHS of \cref{eq:real_stuff} is bounded as follows:
 \begin{align}
 	\abs{\sum\limits_{j=1}^{m} \text{Re} \left\{ \frac{\tr \left(\Delta_{m+1} \Delta_j \right) }{d} \right\} \prod_{\substack{i=1 \\ i \neq j}}^{m} \sqrt{\Phi(\mc D_i^\star, \mc I)}} 
 	&\leq  \sum\limits_{j=1}^{m}	\abs{\text{Re} \left\{ \frac{\tr \left(\Delta_{m+1} \Delta_j \right) }{d} \right\} \prod_{\substack{i=1 \\ i \neq j}}^{m} \sqrt{\Phi(\mc D_i^\star, \mc I)}} \tag{Triangle ineq.} \\
	&\leq  \sum\limits_{i=1}^{m}	\abs{\text{Re} \left\{ \frac{\tr \left( \Delta_{m+1} \Delta_i \right)}{d} \right\} }   \tag{$\sqrt{\Phi(\mc A_i^\star, \mc I)} \leq 1$} \\
	&\leq  \sum\limits_{i=1}^{m}	\abs{\tr \left[ \frac{\Delta_i (\Delta_{m+1} +\Delta_{m+1}^\dagger)/2 }{d} \right] }  \tag{$\Delta_i =\Delta_i^\dagger$ for $i\neq m+1$.} \\
	&\leq  \sum\limits_{i=1}^{m}	\frac{\|\Delta_i\|_2}{\sqrt d} \frac{\|\text{Re}(V)- \tr(V) \mbb I\|_2}{\sqrt d}   \tag{Cauchy-Schwarz ineq.} \\
	\end{align}
	This is where \cref{defn:wse} (equability) is put to use. Recall that for $i \neq m+1$ we have $\frac{\|\Delta_i\|_2}{\sqrt{d}} = \gamma_{\rm decoh}(\mc D_i) \left(1-\sqrt{\Phi(\mc D_i^\star, \mc I)}\right)$ and that the WSE coherence constant is implicitly defined by $\frac{\|\text{Re}(V)- \tr(V) \mbb I\|_2}{\sqrt d}=\gamma_{\rm coh} (1-\sqrt{\Phi(\mc V, \mc I)})$, which means that
	\begin{align}
	\abs{\sum\limits_{j=1}^{m} \text{Re} \left\{ \frac{\tr \left(\Delta_{m+1} \Delta_j \right)}{d} \right\} \prod_{\substack{i=1 \\ i \neq j}}^{m} \sqrt{\Phi(\mc D_i^\star, \mc I)}} 	&\leq  {\gamma_{\rm decoh}}{\gamma_{\rm coh}}\left(1-\sqrt{\Phi(\mc V, \mc I)}\right)\sum\limits_{i=1}^{m}	\left(1-\sqrt{\Phi(\mc{D}_i^\star, \mc I)}\right) \tag{\Cref{defn:wse}} 
\end{align}
Using $|a^2-b^2|\leq |a-b| |a+b|$, and reuniting the pieces, we get
\begin{align}
	\scalemath{0.9}{\abs{\Phi(\mc{V}\circ \mc{D}_{m:1}^\star, \mc I)-\Phi(\mc V, \mc I) \prod_{i=1}^{m} {\Phi(\mc D_{i}^\star, \mc I)}}} &\scalemath{0.9}{\leq 
	2{\gamma_{\rm decoh}}{\gamma_{\rm coh}}\left(1-\sqrt{\Phi(\mc V, \mc I)}\right)\sum\limits_{i=1}^{m}	\left(1-\sqrt{\Phi(\mc{D}_i^\star, \mc I)}\right)}\notag \\
	&\scalemath{0.9}{+ {\gamma_{\rm decoh}^2}\left(\sum\limits_{i=1}^{m} \left(1-\sqrt{\Phi(\mc D_i^\star, \mc I)}\right) \right)^2}\notag \\
	&\scalemath{0.9}{+\abs{\text{Im}\left\{\frac{\tr \left(V |D^m_1|\cdots|D_1^1| \right)}{d}\right\}}^2}~. 
\end{align}
A straightforward application of \cref{eq:star_mult} and \cref{thm:fid_evo} on the LHS (to get rid of the $\star$) yields \cref{eq:quasi_mult_U}.
\end{proof}

\subsection{Proof of \cref{thm:max_cor_multi}}
\begin{proof}
	Let's factor the decoherent channels 
	into their (left) polar decomposition \mbox{$\mc A_i= \mc D_i \circ \mc V_i$}. By relabeling
	$(\mc V_{i:1})^{-1} \circ \mc D_i \circ \mc V_{i:1} = \mc D'_i$ (notice that $\mc D'_i$ are decoherent), 
	we have
	\begin{align}
	\mc A_{m:1}= \mc V_{m:1} \circ \mc {D'}_{m:1}~.
	\end{align}
	First, let's find a lower bound on $\max\limits_{W \in SU(d)} \Phi(\mc W \circ \mc A_{m:1}, \mc U_{m:1})$. A way to do this is to 
	pick a wisely chosen argument for $\mc W$. Let's pick $\mc W = \mc U_{m:1} \circ (\mc V_{m:1})^\dagger $ :
	\begin{align}
	\max\limits_{W \in SU(d)} \Phi(\mc W \circ \mc A_{m:1}, \mc U_{m:1}) 
	&\geq \Phi(\mc U_{m:1} \circ (\mc V_{m:1})^\dagger \circ \mc A_{m:1}, \mc U_{m:1}) \\
	&\geq  \Phi(\mc U_{m:1} \circ (\mc V_{m:1})^\dagger \circ \mc A_{m:1}^\star, \mc U_{m:1}) \tag{\Cref{thm:uni_evo}} \\
	& = \Phi(\mc {D'}_{m:1}^\star, \mc I) \\
	& \geq \prod_{i=1}^m\Phi(\mc {D'}^\star_i,\mc I)+ \gamma_{\rm decoh}^2 \prod_{i=1}^{m} \sqrt{\Phi(\mc D_i^\star,\mc I)} \left(\sum\limits_{i=1}^{m} \left(1-\sqrt{\Phi(\mc D_i^\star, \mc I)}\right) \right)^2
	\notag \\
	&+ ~ \frac{\gamma_{\rm decoh}^4}{4}\left(\sum\limits_{i=1}^{m} \left(1-\sqrt{\Phi(\mc D_i^\star, \mc I)}\right) \right)^4~ \tag{\Cref{eq:star_phi}} 
	\end{align}
	To bound $\prod_{i=1}^{m} \Phi(\mc {D'}_{i}^\star, \mc I)$, we express it as a sum of three terms:
	\begin{align}
	\prod_{i=1}^{m} {\Upsilon(\mc {A}_{i})}+  \left(\prod_{i=1}^{m} {\Upsilon(\mc {A}_{i}^\star)}- \prod_{i=1}^{m} {\Upsilon(\mc {A}_{i})} \right)+\left( \prod_{i=1}^{m} \Phi(\mc {D'}_{i}^\star, \mc I)-\prod_{i=1}^{m} {\Upsilon(\mc {A}_{i}^\star)} \right)~. \label{eq:expansion}
	\end{align}
	To bound the second term, we used \cref{eq:bound_products}.
	The third term of \cref{eq:expansion} is bounded through \cref{eq:telescope}.
	Reuniting the pieces together, we get
	\begin{align}
	\max\limits_{W \in SU(d)} \Phi(\mc W \circ \mc A_{m:1}, \mc U_{m:1})  &\geq 
	\prod_{i=1}^{m} {\Upsilon(\mc {A}_{i})}  - \gamma_{\rm decoh}^2 \sum_{i=1}^{m} \left(1-\sqrt{\Phi(\mc {D}_i^\star,\mc I)}\right)^2
	-\sum_{i=1}^{m} (1-\Upsilon(\mc {A}^\star_{i}))^2 \notag \\
	&- \gamma_{\rm decoh}^2 \prod_{i=1}^{m} \sqrt{\Phi(\mc D_i^\star,\mc I)} \left(\sum\limits_{i=1}^{m} \left(1-\sqrt{\Phi(\mc D_i^\star, \mc I)}\right) \right)^2
	\notag \\
	&- ~ \frac{\gamma_{\rm decoh}^4}{4}\left(\sum\limits_{i=1}^{m} \left(1-\sqrt{\Phi(\mc D_i^\star, \mc I)}\right) \right)^4~.    \label{eq:lower_max}
	\end{align}
	With regards to the upper bound, we can first use \cref{thm:fid_evo} to get
	\begin{align}
	\max\limits_{W \in SU(d)} \Phi(\mc W \circ \mc A_{m:1}, \mc U_{m:1})  &\leq 
	\max\limits_{W \in SU(d)} \Bigg[\Phi(\mc W \circ \mc A_{m:1}^\star, \mc U_{m:1})+ \frac{1}{2} \left(\sum\limits_{i=1}^m (1-\Upsilon(\mc A_i^\star))\right)^2 \notag \\
	&~~+(1- \Phi(\mc W \circ \mc A^\star_{m:1}, \mc U_{m:1}))\sum \limits_{\substack{i=1} }^{m}\left(1- \Upsilon(\mc A_i^\star)\right) \Bigg]~.\label{eq:up_max}
	\end{align}
	By using the flavored Von-Neumann inequality (\cref{lem:vn_mat_ineq}), followed by \cref{eq:uni_fid_id}, we get 
	\begin{align}
	\max\limits_{W \in SU(d)} \Phi(\mc W \circ \mc A_{m:1}^\star, \mc U_{m:1}) &\leq 
	\abs{\frac{\tr |D'_{m:1}|}{d}}^2 \tag{\Cref{lem:vn_mat_ineq}}\\
	& \leq {\Upsilon(\mc {D'}_{m:1}^\star)} \tag{\Cref{eq:uni_fid_id}} \\
	& \leq \prod_i {\Upsilon(\mc {A}_i^\star)} + 2 \gamma_{\rm decoh}^2 \left(\sum_{i=1}^m \left(1-\sqrt{\Phi(\mc {D}_i^\star,\mc I)}\right)\right)^2~, \tag{\Cref{eq:uni_before_sq}} \\
	& \leq \prod_i {\Upsilon(\mc {A}_i)} + \sum_{i=1}^{m} (1-\Upsilon(\mc {A}^\star_{i}))^2 \notag \\
	& ~~+ 2 \gamma_{\rm decoh}^2 \left(\sum_{i=1}^m \left(1-\sqrt{\Phi(\mc {D}_i^\star,\mc I)}\right)\right)^2~. \tag{\Cref{eq:bound_products}}
	\end{align}
	Substituting this on the RHS of \cref{eq:up_max} completes the proof.
\end{proof}    

\end{document}

%% file: macros.tex
\usepackage{hyperref,graphicx, amsmath, amssymb, amsthm, color, bm, bbm,dsfont, cleveref,media9,doi,floatrow,enumerate}
\usepackage[most]{tcolorbox}

\newtcbtheorem{boxthm}{Theorem}{
	enhanced,
	sharp corners,
	attach boxed title to top left={yshift=-\tcboxedtitleheight/2, xshift=3mm,yshifttext=-\tcboxedtitleheight/2},
	colback=white,
	colframe=red!20,
	coltitle=black,
	boxrule=0.7mm,
	fonttitle=\bfseries,
	boxed title style={
		sharp corners,
		size=small,
		colback=red!20,
		colframe=red!20,
	} 
}{thm}
\crefname{tcb@cnt@boxthm}{theorem}{theorems}

\newtcbtheorem{boxlem}{Lemma}{
	enhanced,
	sharp corners,
	attach boxed title to top left={yshift=-\tcboxedtitleheight/2, xshift=3mm,yshifttext=-\tcboxedtitleheight/2},
	colback=white,
	colframe=blue!20,
	coltitle=black,
	boxrule=0.7mm,
	fonttitle=\bfseries,
	boxed title style={
		sharp corners,
		size=small,
		colback=blue!20,
		colframe=blue!20,
	} 
}{lem}
\crefname{tcb@cnt@boxlem}{lemma}{lemmas}

\newtcbtheorem{boxdefn}{Definition}{
	enhanced,
	sharp corners,
	attach boxed title to top left={yshift=-\tcboxedtitleheight/2, xshift=3mm,yshifttext=-\tcboxedtitleheight/2},
	colback=white,
	colframe=green!55!black!15,
	coltitle=black,
	boxrule=0.7mm,
	fonttitle=\bfseries,
	boxed title style={
		sharp corners,
		size=small,
		colback=green!55!black!15,
		colframe=green!55!black!15,
	} 
}{defn}
\crefname{tcb@cnt@boxdefn}{definition}{definitions}

\usepackage[framemethod=TikZ]{mdframed}
\usepackage[most]{tcolorbox}
\theoremstyle{plain}

\theoremstyle{definition}

\crefname{thm}{theorem}{theorems}

\newcounter{theo}

\crefname{theo}{theorem}{theorems}

\newcounter{lembx}

\crefname{lembx}{lemma}{lemmas}

\newcounter{defbx}

\crefname{defbx}{definition}{definitions}

\definecolor{nblue}{rgb}{0.2,0.2,0.7}
\definecolor{ngreen}{rgb}{0.1,0.5,0.1}
\definecolor{nred}{rgb}{0.8,0.2,0.2}
\definecolor{nblack}{rgb}{0,0,0}

\DeclareMathOperator{\tr}{\mathrm{Tr}}
\usepackage{scalerel}
\usepackage{stackengine,wasysym}

\newcommand{\mc}[1]{\mathcal{#1}}

\newcommand{\mbb}[1]{\mathbb{#1}}

\newcommand{\inner}[2]{\left\langle#1,#2\right\rangle}
\newcommand{\abs}[1]{\left| #1 \right|}
\newcommand{\nlz}[1]{\underbar{$#1$}}

\newcommand{\hidden}[1]{}

\newcommand{\fontten}{\fontsize{10pt}{12pt}\selectfont}

\newcommand\scalemath[2]{\scalebox{#1}{\mbox{\ensuremath{\displaystyle #2}}}}